\documentclass[aip,pop,sd,reprint]{revtex4-1}

\usepackage{amsmath,framed,amsthm}
\usepackage{amssymb,amsfonts,amstext,graphics,graphicx,subfigure}
\usepackage[colorlinks]{hyperref}
\usepackage{color}
\usepackage{mathbbol}

\DeclareSymbolFontAlphabet{\amsmathbb}{AMSb}%
 
\def\XXint#1#2#3{{\setbox0=\hbox{$#1{#2#3}{\int}$ }
\vcenter{\hbox{$#2#3$ }}\kern-.5\wd0}}

\def\cale{\mathcal{E}}
\def\calf{\mathcal{F}}

\def\calj{\mathcal{J}}
\def\calk{\mathcal{K}}

\def\cals{\mathcal{S}}

\def\G{\mathbb{G}}

\def\C{\mathbb{C}}
\def\R{\mathbb{R}}
\def\C{\mathbb{C}}

\def\bq{\begin{equation}}
\def\eq{\end{equation}}
\def\bqy{\begin{eqnarray}}
\def\eqy{\end{eqnarray}}

\def\al{\alpha}
\def\be{\beta}

\def\de{\delta}
\def\De{\Delta}
\def\ep{\epsilon}

\def\ga{\gamma}

\def\om{\omega}
\def\Om{\Omega}

\def\ps{\psi}

\def\si{\sigma}

\def\th{\theta}

\def\ze{\zeta}

\def\bfA{\mathbf{A}}
\def\bfB{\mathbf{B}}

\def\bfE{\mathbf{E}}
\def\bfa{\mathbf{a}}
\def\bfJ{\mathbf{J}}
\def\bfk{\mathbf{k}}

\def\bfx{\mathbf{x}}

\def\bfv{\mathbf{v}}

\def\bfq{\mathbf{q}}

\def\bfz{\mathbf{z}}

\def\ringz{\mathring{z}}
\def\ringx{\mathring{x}}
\def\ringv{\mathring{v}}

\def\p{\partial}

\def\and{\quad\mathrm{and}\quad}

\def\Brac#1#2{\{#1,#2\}}
\def\brac#1#2{[#1,#2]}

\DeclareMathAlphabet{\mathpzc}{OT1}{pzc}{m}{it}

\newcommand{\jac}{\ensuremath{\mathcal{J}}}

\newcommand{\dd}{\ensuremath{\mathrm{d}}}

\newcommand{\Dt}{\ensuremath{\Delta t}}

\renewcommand{\vec}[1]{\ensuremath{\mathbf{#1}}}
\newcommand{\coeff}[1]{\ensuremath{\lowercase{#1}}}

\newcommand{\X}{\ensuremath{\vec{X}}}
\newcommand{\V}{\ensuremath{\vec{V}}}

\newcommand{\ca}{\ensuremath{\vec{\coeff{A}}}}
\newcommand{\cb}{\ensuremath{\vec{\coeff{B}}}}

\newcommand{\ce}{\ensuremath{\vec{\coeff{E}}}}
\newcommand{\cj}{\ensuremath{\vec{\coeff{J}}}}
\newcommand{\cphi}{\ensuremath{\boldsymbol{\varphi}}}
\newcommand{\crho}{\ensuremath{\boldsymbol{\varrho}}}

\newcommand{\MM}{\ensuremath{\mathbb{M}}}

\newcommand{\D}{\ensuremath{\mathbb{D}}}

\newcommand{\Lab}{\ensuremath{\boldsymbol{\Lambda}}}
\newcommand{\LaB}{\ensuremath{\mathbb{\Lambda}}}

\begin{document}

\title{Structure and structure-preserving algorithms for plasma physics}

\author{P. J. Morrison}

\email[Electronic Mail: ]{morrison@physics.utexas.edu}
\homepage[\\ \hspace*{.25 cm}Homepage: ]{https://web2.ph.utexas.edu/~morrison/}

\affiliation{Department of Physics and Institute for Fusion Studies, 
The University of Texas at Austin, Austin, TX, 78712, USA}

\date{\today}

\begin{abstract}
Hamiltonian and action principle (HAP) formulations of plasma physics are reviewed for the purpose of explaining structure preserving numerical algorithms.  Geometric structures associated with and emergent from HAP formulations are discussed.  These include conservative integration, which exactly conserves invariants, symplectic integration, which exactly preserves the Hamiltonian geometric structure, and other Hamiltonian integration techniques. Basic ideas of variational integration and Poisson integration, which can preserve  noncanonical Hamiltonian structure,  are discussed.  Metriplectic integration, which preserves the structure of conservative systems with  both Hamiltonian and dissipative parts,  is proposed. Two kinds of simulated annealing, a relaxation technique for obtaining equilibrium states, are  reviewed:  one that uses  metriplectic dynamics, which maximizes an entropy at fixed energy,  the other that uses double bracket  dynamics, which preserves Casimir invariants.  Throughout,  applications to plasma systems are emphasized.  The paper culminates with  a discussion of GEMPIC [Kraus et al.,  arXiv:1609.03053v1 [math.NA] (2016)], a particle in cell code that incorporates Hamiltonian and geometrical structure preserving properties.

\medskip

\end{abstract}

\maketitle

 
\section{Introduction}
\label{sec:intro}

The purpose of this article is to give an overview of computational algorithms designed to preserve structure of dynamical systems,  and to review some recent progress in the development of structure preserving  algorithms for plasma simulation.    

The first question, of course, is what is meant by structure.  Clearly all computational scientists  attempt  to preserve the solution of their systems, the ultimate structure, which would mean this article would be about the proud history of many decades of computational plasma physics research that attempts  to do just that.    Thus it is necessary to rein in the purview, and this is done by concentrating  on properties associated with or that emerge from the Hamiltonian and action principle (HAP) structure of plasma models, structure that ultimately originates  from  the fundamental electromagnetic interaction.   Of interest are algorithms  of relevance to plasma science that preserve various HAP properties that occur in finite systems, ordinary differential equations (ODEs), e.g., those  that describe  particle orbits or magnetic field lines,  and  in  infinite systems of partial differential equations (PDEs),  plasma field theories,  e.g.,   various plasma fluid models and  kinetic theories.  Clearly ever  since the advent of  computation,  researchers have attempted to design methods of structure preservation (see e.g.\ Refs.~\onlinecite{vogelaere,dewitt53}  for early examples), and there has been continual progress over the years (see e.g.\  Refs.~\onlinecite{Mclachlan01,HairerLubichWanner:2006} for more modern surveys).   The perspective  taken here is to describe in broad brush strokes the types of structure that can be preserved, illustrated with selected examples; however,  global value judgements and comparisons  of the efficacy of various  techniques will be avoided.  Many references are cited, but constructing a complete bibliography would be a daunting if not impossible task,  even with the given limitation of scope.  Rather,  a  representative although somewhat biased selection of literature  is  given,  but one that is hoped will provide a gateway  into the field.

 The article also has the goal of highlighting the community of plasma researchers  working in this area, as exemplified by workshops organized by Hong Qin, Eric  Sonnendr\"ucker, and myself:
\begin{itemize}
\item Geometric Algorithms and Methods for Plasma Physics Workshop (GAMPP) 
May 13--15, 2014 Hefei, China (Hong Qin)
\item  GAMPP II, September 12--16, 2016 in Garching, Germany (PJM, Hong Qin, and  Eric Sonnendr\"ucker)
\item  Mini-Conference at this meeting: New Developments in Algorithms and Verification of Gyrokinetic Simulations (Amitava Bhattacharjee and Eric Sonnendr\"ucker)
  \end{itemize}
  It is possible there will be a  GAMPP III in  2018.
 
 \medskip

The HAP structure of classical plasma physics  originates from the relativistic action principle for  a collection of charged particles interacting self-consistently with the electromagnetic fields they generate.  This system has  the dynamical variables $(\bfq_{i}(t), \phi(\bfx,t),  \bfA(\bfx,t))$, where $\bfq_{i}$ is the position of the $i$th particle, $\phi$ is the electrostatic potential, and $\bfA$ the vector potential.  The Lorentz covariant action is given by
\begin{eqnarray} 
 S[\bfq,\phi,\bfA] &=& - \sum_s\sum_{i=1}^{N}\int\! dt \,
     \ mc^{2} \sqrt{1 - \frac{|{\dot{\mathbf{q}}_{i}}|^{2}}{c^{2}}}\, 
      \label{kinetic}\\
         && \hspace{-.6 cm}  - \sum_s e \int \!dt\sum_{i=1}^{N} \, \int \!d^3x\,   
     \Big[\phi(\bfx,t) 
   \nonumber  \\
     && \hspace{1 cm} +\  \frac{{\dot{\mathbf{q}}_{i}}}{c} \cdot \mathbf{A}(\bfx,t) \Big]\,
     \delta\left(\bfx-\bfq_{i}(t)\right)
    \label{coupling}\\
    &&  \hspace{-.6 cm} +\  \frac1{8\pi}\int \!dt\! \int \!d^3x \,\Big[|\mathbf{E}(\bfx,t) |^{2}- 
|\mathbf{B}(\bfx,t)|^{2}\Big]   ,
\label{field}
\end{eqnarray}
where \eqref{kinetic} represents the  relativistic  kinetic  energy,    \eqref{coupling} provides the   coupling between the particle dynamics and the fields, and \eqref{field} is the pure  field contribution.  Here  $\sum_s$ is a sum over species, $N$ is the number of particles of each species,  and  species indices on $N$, $\bfq_i$, $e$, and  $m$,  are suppressed to avoid clutter.  In \eqref{field} $\bfE$ and $\bfB$ are shorthand notation  for the usual expressions in terms of the scalar and vector potentials 
\bq
\bfE = -\nabla \phi  - \frac1{c} \frac{\p \bfA}{\p t} \quad \mathrm{and}\quad \bfB=\nabla\times \bfA\,.
\label{potentials}
\eq
Faraday's law and $\nabla\cdot \bfB=0$ follow automatically from these expressions.

Extremization of \eqref{kinetic}--\eqref{field} yields Maxwell's equations coupled to particle dynamics.
We will do this extremization by the equivalent procedure of functional differentiation, which we carefully  define here because it will be used extensively  later on.  For the variation of $S$ with respect to $\phi$ we consider the functional derivative $\de S/\de \phi$, which  is defined as follows: 
\bq
 \left.\frac{d}{d\ep}S[\bfq, \phi + \ep \de\phi, \bfA]\right|_{\ep=0}
\hspace{-.3cm}= \int \!dt\! \int \!d^3x\,   \de  \phi(\bfx,t)\,  \frac{\delta S}{\delta \phi(\bfx,t)} 
\label{functD}
\eq
where $\de \phi$ is an arbitrary function of its arguments,  subject to admissible boundary conditions that assure the vanishing of terms obtained by integrations by parts.  In \eqref{functD}, after $\de\phi$ is isolated  one can extract  ${\delta S}/{\delta \phi}(\bfx,t)=0$ by removing the integrals and $\de\phi$.  The argument $(\bfx,t)$ of  $\de S/\de \phi(\bfx,t)$ is displayed so one knows which integrals to remove.  (See Ref.~\onlinecite{pjm98} for an extended discussion of functional differentiation.) For the action of 
\eqref{kinetic}--\eqref{field}  the functional derivative ${\delta S}/{\delta \phi}$ defined by \eqref{functD} yields  Poisson's equation 
\bq
\nabla\cdot\bfE=4\pi\rho= \sum_s\sum_i e\, \de(\bfq_i-\bfx)\,, 
\eq
as expected.  Similarly,  the Amp\`ere-Maxwell law follows from from $\de S/\de \bfA(\bfx,t)$,
 \bq
 \frac{\p \bfE}{\p t}= c \nabla\times\bfB - 4 \pi \bfJ
 \eq
 where  the  current density is  given  by
\bq
\bfJ(\bfx,t) = \sum_s \sum_i e\, \dot{\bfq}_i\, \de(\bfq_i-\bfx)\,.
\label{current}
\eq
Lastly, the variation on particles is given by $\de S/\de \bfq_i(t)$, and this produces  the relativistic version of Newton's second law.  This  functional derivative  involves integrating by parts in  time and performing the  integration over $d^3x$ so as to evaluate the Lorentz force on the particle positions.  Thus the coupling term provides both the sources in Maxwell's equations and the force for  the particle dynamics.   
 
 The lion's share of plasma physics is embodied in the action principle of \eqref{kinetic}--\eqref{field}, so from one lofty point of view the discipline  is complete.  However, this naive point of view misses the beautiful emergent collective phenomena of plasma science that has required considerable effort to unravel.  The unraveling has entailed  various approximations, reductions resulting in various fluid and kinetic models  that elucidate  various plasma phenomena.  Generally speaking, the reduced models that originate from  the action $S$  contain both dissipative and nondissipative processes, and the action principle origin can be obscured. Nevertheless, HAP properties bubble to the surface as part of  essentially  all  plasma models,  ranging from magnetic field line behavior, to MHD and more general magnetofluid dynamics, to various kinetic theories, and even the BBGKY hierarchy itself.  We refer the reader to Refs.~\onlinecite{pjm82,pjmY92,pjm98,pjm05,pjm06,pjm09,pjmKLWW14,pjmAP16,pjmKM16},  which are reviews or contain a significant review component of HAP structure.

The paper is organized as follows.   Section~\ref{sec:CI} discusses  conservative integration, schemes that are designed to exactly conserve constants of motion.  Next,     Sec.~\ref{sec:SI} treats  symplectic integrators that preserve canonical Hamiltonian structure, which is followed by   Sec.~\ref{sec:HI} where  various other forms of canonical and noncanonical Hamiltonian integration are described, including variational integration and Poisson integration.   Section \ref{sec:MI}  describes  metriplectic  integration, a proposal to build structure preserving  integrators for systems that are conservative with  both Hamiltonian and dissipative parts.  Simulated annealing, a relaxation method that preserves structure while  obtaining equilibrium states,  is considered in Sec.~\ref{sec:SA}.   In Sec.~\ref{sec:gempic}  GEMPIC algorithms for the Maxwell-Vlasov (MV) system are described.  These are particle in cell (PIC) codes that preserve geometric structure, including the noncanonical Hamiltonian structure of the MV system,  its Casimir invariants,  and the geometry of Maxwell's equations.    Here many of the features of previous sections are incorporated: it  is shown how to obtain a  semidiscrete  Hamiltonian reduction of the MV system,  yielding a  finite-dimensional  Hamiltonian system with noncanonical Poisson, which is then integrated with a Poisson integrator.   Finally, the paper concludes with Sec.~\ref{sec:conclu}.

\section{Conservative Integration (CI)}
  \label{sec:CI}
  
Generally speaking, reductions of the system with the HAP formalism of the action of \eqref{kinetic}--\eqref{field} (and associated Hamiltonian structure)  based on time scale separation result in non-HAP reductions. Examples of this kind of reduction include truncation of the BBGKY hierarchy leading, to  the Landau-like collision operators,  and  the derivation of quasilinear theory.  For these models the Hamiltonian structure is lost, but conservation of energy  and possibly other invariants,  may be  maintained.   Sometimes one is aware that energy survives an approximation, by direct calculations after the fact or by construction, or that an unknown  Hamiltonian structure actually exists. 
For some such systems, the long time  states may be governed to a large degree by  the preservation of a few invariants, whether or not the system is Hamiltonian.  

The failure of an integrator  to preserve an invariant  is demonstrated by the phenomenon of  energy drift.  To understand this consider a system with a conserved energy or part of an energy given by a kinetic energy expression  of the form
\[
E= m |\mathbf{v}|^2/2 \,,
\]
where $m$ is a particle mass and $\bfv$ its velocity. Let $\mathbf{v}_{\mathrm{calculated}}$ be the value of the velocity produced by some algorithm and $\de \mathbf{v}$ the error, both  after a single time step. Thus  
\[
\mathbf{v}_{\mathrm{calculated}}= \mathbf{v}_{\mathrm{exact}} + \de \mathbf{v}
\]
and after a single time step 
\bqy
E_{\mathrm{calculated}}&=&  m\,  |\mathbf{v}_{\mathrm{exact}}+  \de \mathbf{v} |^2/2
\nonumber\\
&=&
m\, |\mathbf{v}_{\mathrm{actual}}|^2/2 
+   m\, \mathbf{v}_{\mathrm{exact}}\cdot \de \mathbf{v} +  m\, |\de \mathbf{v} |^2 /2\,.
\nonumber
\eqy
Over many time steps one may expect the cross term $\mathbf{v}_{\mathrm{exact}}\cdot \de \mathbf{v}$ to average to zero, while the term $|\de \mathbf{v} |^2$ will systematically give rise to a monotonically changing $E$. 
Indeed,  for some standard integrators,  energy may tend to increase dramatically. 

Over the years many researchers have used various techniques for enforcing the  conservation of invariants.  If one begins with a PDE then a common mode of development is  to first obtain a semidiscrete approximation, an energy conserving set of differential equations, and then a next step would be to construct an integrator that enforces energy conservation.  There are many examples  of this is procedure; we mention spectral reduction\cite{pjmBS99} that was designed to give accurate spectral and cascades for turbulence described by the 2D Euler equation.  For this system a semidiscrete reduction that exactly conserves energy and enstrophy was obtained, as desired for accurate spectra, and the resulting set of ordinary differential equations was then solved by a conservative integrator.  Other examples include Refs.~\onlinecite{scott08,scott13}, pertaining to  turbulent transport of relevance to magnetic confinement,  and Refs.~\onlinecite{chacon,chenglong1,chenglong2} that consider Fokker-Planck type collison operators.  In Ref.~\onlinecite{chacon}  two species interact through collisions and exact energy conservation is required for proper thermalization of the two species.  Both of these systems are examples where one seeks long time behavior and that behavior is sensitive   to conservation laws.   Many examples of  conservative or  nearly conservative integration exist including, e.g.,  one based on Hermite expansion in Ref.~\onlinecite{delzano15},  the discontinuous Galerkin method, RKDG,  developed in Ref.~\onlinecite{shu89} and used in the plasma context  for  the Vlasov equation in Refs.~\onlinecite{pjmCG13,pjmCGL14},  and at this meeting in Ref.~\onlinecite{hammettAPS}.

\subsection{Conservation Structure -- Sets of Invariants}
\label{ssec:CIS}

A set of $M$ first order ordinary differential  equations (ODEs), written as
 \bq
  \dot{z}^a=V^a(z) \qquad   a=1,2,\dots, M\,, 
  \label{Code}
  \eq
where dot denotes total time differentiation and  $V$ is some vector field, possesses  a conserved  energy function $E(z)$ provided 
\bq
  \dot E= \frac{\p E}{\p z^a}\,\dot z^a= V^a \,  \frac{\p E}{\p z^a}= 0\,,
  \label{dotE}
 \eq
where in the first equality of \eqref{dotE}, which follows from the chain rule,  repeated summation over the index $a$ is assumed.   For such systems
\bq
E (z(t) ) = E (\mathring{z})\,,
\label{Eode}
\eq
with $\mathring{z}$ being the initial value of the dependent variable $z=(z^1,z^2, \dots, z^M)$.  Similarly, any additional  invariant $I$ must satisfy $V^i  \p I/ \p z^i= 0$.

For a partial (integro) differential equation (PDE) we write 
\bq
  \chi_t= \mathcal{V}(\chi,\chi_\mu,\dots,)\,,
  \label{Cpde}
 \eq
where $\chi(\mu,t)$ is a multicomponent field variable dependent on a `spatial' variable $\mu$ of any dimension and time. For such PDEs,  subscripts will denote partial differentiation.

Systems of the form of \eqref{Cpde} possess an invariant  energy functional $E[\chi]$ if  
 \bq
  \dot{E}[\chi]=\int \!d\mu \, \frac{\de E}{\de \chi}\cdot \chi_t 
  = \int \!d\mu\,  \mathcal{V}\cdot \frac{\de E}{\de \chi} =0\,,
 \label{dotEpde}
\eq
where  $\de E/\de \chi(\mu)$ denotes the functional derivative as described in Sec.~\ref{sec:intro}, here defined in terms of the pairing
\bq
\de E[\chi;\de\chi]=\left.\frac{d}{d\ep} E[\chi + \ep \de \chi]\right|_{\ep=0}\!\!\!
=\int\!d\mu\, \frac{\de E}{\de \chi}\cdot \de\chi\,,
\eq
 where the first variation $\de E[\chi;\de\chi]$, a linear operation on $\de\chi$, is the Fr\'echet derivative.  The `$\,
\cdot\,$' of \eqref{dotEpde} is a shorthand for summation over  the components of the field $\chi$ should it have more  than one.  The goal of a CI is to have the following exactly satisfied:
\bq
E[\chi]=E[\mathring{\chi}] 
\label{Epde}
\eq
with $\mathring{\chi}$ being the initial value of $\chi$.  

Thus, the goal of a conservative integrator is to construct a numerical algorithm to maintain exactly 
\eqref{Eode} for an ODE or \eqref{Epde} for a PDE. Or, if there are other known invariants, to also preserve these as desired.

\subsection{Conservative Integrators}
\label{ssec:CI}
  
Although various means have been used to preserve invariants, a systematic procedure amenable to various types of integrators was developed and applied in Refs.~\onlinecite{pjmSB99,pjmBS99,bowman}.  The procedure of these references can be used to adapt common integrators, such as Euler's method, predictor-corrector, etc.,  so as to  exactly conserve an invariant or a family of invariants. 
  
The procedure of Ref.~\onlinecite{pjmSB99} is quite simple.  For a set of ODEs, first one modifies the system by adding a `correction' vector field $V_C$ to \eqref{Code}, i.e.,  this system becomes 
\bq
\dot{z}= V + V_C\,.
\eq
Next, it is required that 
\bq
  \lim_{\De t\rightarrow 0}  V_C =  0
 \eq
 and this must be at a rate fast enough so as to preserve the order of the desired integrator. 
 Finally, one chooses $V_C$ so has to enforce 
\bq
E\big(z(t + {\De t})\big) = E\big(z(t) \big)
\label{ECI}
\eq
exactly  at each time step.  When  \eqref{ECI} is only a single  constraint for a system of equations,  there is freedom on how to enforce this;  if a family of invariants exists, then one has more equations like  \eqref{ECI}, but there is still latitude for enforcing the family.

Several physical examples are solved with  this type of CI  in Refs.~\onlinecite{pjmSB99,pjmBS99,bowman}; in particular,  Euler's method is  applied to Euler's rigid body equations,  as a simple proof of principle example.   Another example, one   that illustrates well  the role invariants can play in answering a question  about the long time behavior  of a system is the famous Kepler  problem.  Although,  this problem is integrable, with  an easily evaluated solution,  it serves as a  illustrative example.  The computational question of interest is to obtain  the path traced out by the orbit over a long period of time, a question that is critically dependent on invariance. 

 As usual, the Kepler problem can be  reduced to motion in a plane,  governed by the following three ODEs: 
\bqy
 \dot r&=&v_r
\label{dotr}\\
 \dot v_r&=&\frac{\ell^2}{m^2 r^2} - \frac1{m}\phi'
\label{dotv}\\
\dot \th&=&\frac{\ell^2}{m r^3}
\label{dotth}
\eqy
where  $(r,\th)$ are polar coordinates, the gravitational potential $\phi=-K/r$ with constant $K$, $\ell$ is the magnitude of the angular momentum, and $m$ is the reduced mass.   This system conserves the energy
  \bq
  E= \frac{m v_r^2}{2} + \frac{\ell^2}{2m^2 r^2} +\phi(r) 
\eq
plus an extra invariant,  the Runge-Lenz,  
\bq
 \mathbf{A}= \mathbf{v}\times \mathbf{L} + \mathbf{r} \phi\,.
\eq 
The Runge-Lenz serves to lock together the radial and angular frequencies for all values of the energy -- as a consequence all bounded orbits ($E<0$) are ellipses independent of  chosen  initial conditions.    With the initial conditions $\mathring{r}$, $\mathring{\th}=0$, and $\mathring{v}_r=0$, $\bfA$ only has the $x$-component
\[
A = \frac{\ell^2}{m  \mathring{r}} -K\,.
\]

\begin{figure} 
\centering
\includegraphics[width=7.0cm]{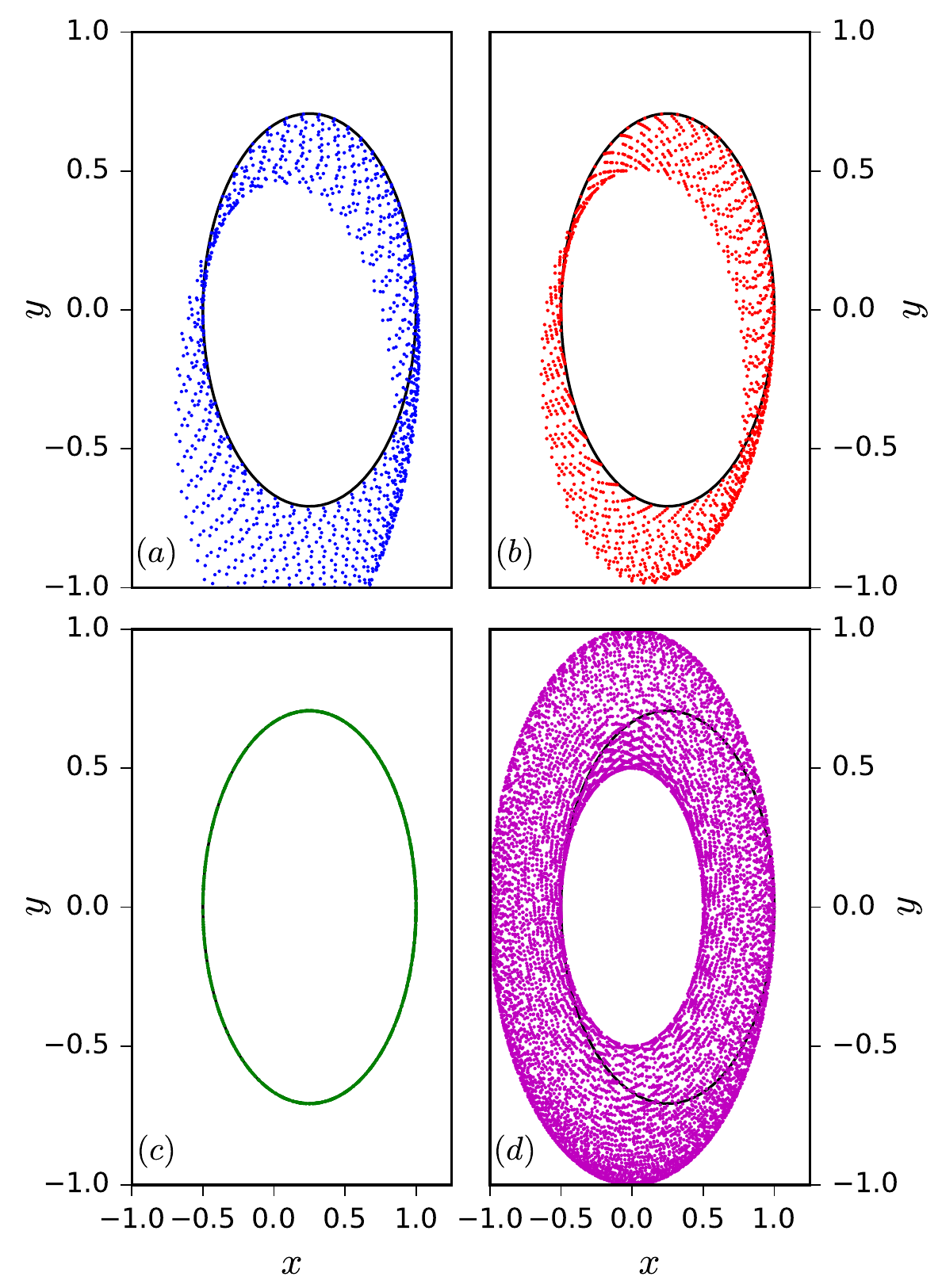}
\caption{Solutions of the Kepler problem (Courtesy of B.\ A.\ Shadwick. See Ref.~\onlinecite{pjmSB99}.) (a) computed using the PC algorithm with a total of 1313 fixed time steps of size 0.08; (b) computed using the PC algorithm with fixed time steps of size 0.08 and only energy conserved; (c) computed using the C-PC algorithm with both energy and Runge-Lenz (RL) vector conserved for a total of 1000 fixed time steps of size 0.105;  (d) long time integration with only energy conserved, showing precession due to violation of the RL conservation.} 
 \label{fig:RL1}
\end{figure}

Figure \ref{fig:RL1} depicts calculated and known orbits for second order predictor corrector (PC) with and without  the CI constraints (see Ref.~\onlinecite{pjmSB99} for details).  In panel (a) plain PC is used and we see the calculated orbit deviate significantly from the known exact elliptical path.  In panel (b) energy is conserved, but not the RL vector resulting in a wobbling precession.  Panel (c) shows the case where both energy and the RL vector are conserved, giving rise to near perfect alignment of the calculated and know path.  In panel (d) we have run the integrator with only energy conservation for a long time depicting  the expected annular region sampled  because of precession.    Similarly Runge's RK4 gives poor performance in conserving Runge's vector. Thus, if one is interested in the planet path, straight integrations may give poor performance, while  a CI can be a valuable tool.

Often physical systems are conservative, yet possess vector fields that have Hamiltonian and non-Hamiltonian components.   For example, this is the case for the Vlasov equation with  the Landau-Fokker-Planck collision operator.   A formalism that describes the nature of this ubiquitous occurrence, called metriplectic dynamics in Ref.~\onlinecite{pjm86}, will be discussed in Sec.~\ref{sec:MI} where metriplectic integrators (MIs) will be proposed.   Before discussing  MIs,  a combination of a CI and a HI, various forms of HIs will be discussed in Secs.~\ref{sec:SI} and \ref{sec:HI}.

  \section{Symplectic Integration (SI)}
  \label{sec:SI}
  
  Whereas the geometric content of CI is fairly minimal, i.e.,  the  preservation of  a few  constraints  restricting   the dynamics to submanifolds, SI is both global and  local in nature and amounts to preserving  the entire symplectic geometry of phase space, a consequence of the Hamiltonian form.  Symplectic geometry   allows one to measure sizes of particular  two-dimensional and higher even dimensional objects,  akin to the measurement of lengths and angles of Riemannian geometry. 

The idea behind SI is old:   a written account occurred in 1956 in a report of  de Vogelaere\cite{vogelaere} and was widely known and used in the accelerator physics community.  In fact, prior to this paper, symplectic algorithms  were used unbeknownst to  their  originators, viz.\  the method used by Verlet and St\"ormer and the leapfrog method.   A  flurry of papers on symplectic integrators appeared in the 1980s by various authors, notably Ref.~\onlinecite{ruth83}.  A general and  interesting historical account  from an  accelerator physics point of view is given in   Ref.~\onlinecite{forest}. 

In the plasma physics community SI in the infinite-dimensional context has been employed, where it has been used to integrate semidiscrete Hamiltonian equations  for describing the two-stream instability\cite{pjmK95,pjmK95b} and for understanding facets of the  nonlinear plasma evolution in the single and multiwave models\cite{pjmTM94,cary0,cary1} of the bump-on-tail instability.

\subsection{Canonical Hamiltonian Structure}
\label{ssec:CHS}
  

The concept of SI is  a consequence of  Hamilton's equations,
\bq
\dot{q}^i=\frac{\p H}{\p p_i}
\quad \mathrm{and}\quad 
\dot{p}^i=-\frac{\p H}{\p q_i}\,, \quad  i=1,2,\dots N\,,
\label{eqHam}
\eq
a set of $2N$ first order differential equations determined entirely by the Hamiltonian function $H(q,p)$ depending on the canonical  coordinates $q$ and momenta $p$.  
Sometimes it is convenient to let  $z=(q,p)$  and rewrite \eqref{eqHam} in tensorial form 
\bq
\dot{z}^{\al}=J_c^{\al\be} \frac{\p H}{\p z^{\be}}=\{ z^\al, H\}\,, \quad \al,\be=1,2,\dots, 2N 
\label{zHam}
\eq
where the canonical Poisson matrix and bracket are given, respectively, by 
\bq
J_c=
\left( 
\begin{array}{cc}
0_N & I_N \\
-I_N & 0_N 
\end{array}
 \right) \quad \mathrm{and}\quad   \{f,g\}=  \frac{\p f}{\p z^{\be}}J_c^{\al\be} \frac{\p g}{\p z^{\be}}\,, 
 \label{CPB}
\eq
with  $0_N$ being an $N\times N$ matrix of zeros,  $ I_N$ being the $N\times N$ identity matrix, and $f$ and $g$ being arbitrary functions of $z$. 

An important  idea of Hamiltonian mechanics is to effect  special coordinate transformations  to obtain solutions or approximate solutions. Upon representing  a coordinate change as $z=z(\bar{z})$,  $2N$ functions of $2N$ variables, with the inverse $\bar{z}=\bar{z}({z})$ Eqs.~\eqref{zHam} become
 \[
 \dot{\bar{z}}^{\al}=\bar{J}^{\al\be} \frac{\p \bar{H}}{\p \bar{z}^{\be}}\quad  {\mathrm{with}} \quad \bar{H}(\bar{z})= H(z)\,,
 \]
 where we see that $H$ transforms as a scalar  and the  Poisson matrix $J$ as a  contravariant 2-tensor, 
 \bq
 \bar{J}^{\al\be}(\bar{z})=  {J_c}^{\mu\nu} \frac{\p \bar{z}^\al}{\p  {z}^{\mu}}\frac{\p \bar{z}^\be}{\p  {z}^{\nu}} = {J_c}^{\al\be}\,.
 \label{Jtrans}
 \eq
If the second equality of \eqref{Jtrans} is satisfied, then the transformation is called a  canonical transformation (CT) (or symplectomorphism in mathematics) and the equations in the new coordinates have clearly identified new coordinates and momenta $\bar{z}=(\bar{q},\bar{p})$ and, consequently,  the form of  Hamilton's equations is preserved:
\[
\dot{\bar{q}}^i=\frac{\p \bar{H}}{\p \bar{p}_i}
\quad \mathrm{and}\quad 
\dot{\bar{p}}_i=-\frac{\p \bar{H}}{\p \bar{q}^i}\,, \quad i=1,2,\dots, N\,.
 \]

A main theorem of Hamiltonian dynamics is that the time advance map is itself a CT; i.e.,   if $z=z(\mathring{z}, t)$ is the solution of  \eqref{zHam} at time $t$ with initial condition $\mathring{z}$, then for any fixed $t$ the coordinate change $z\leftrightarrow \mathring{z}$ is a CT.

\begin{figure} 
\centering
 \includegraphics[width=7 cm]{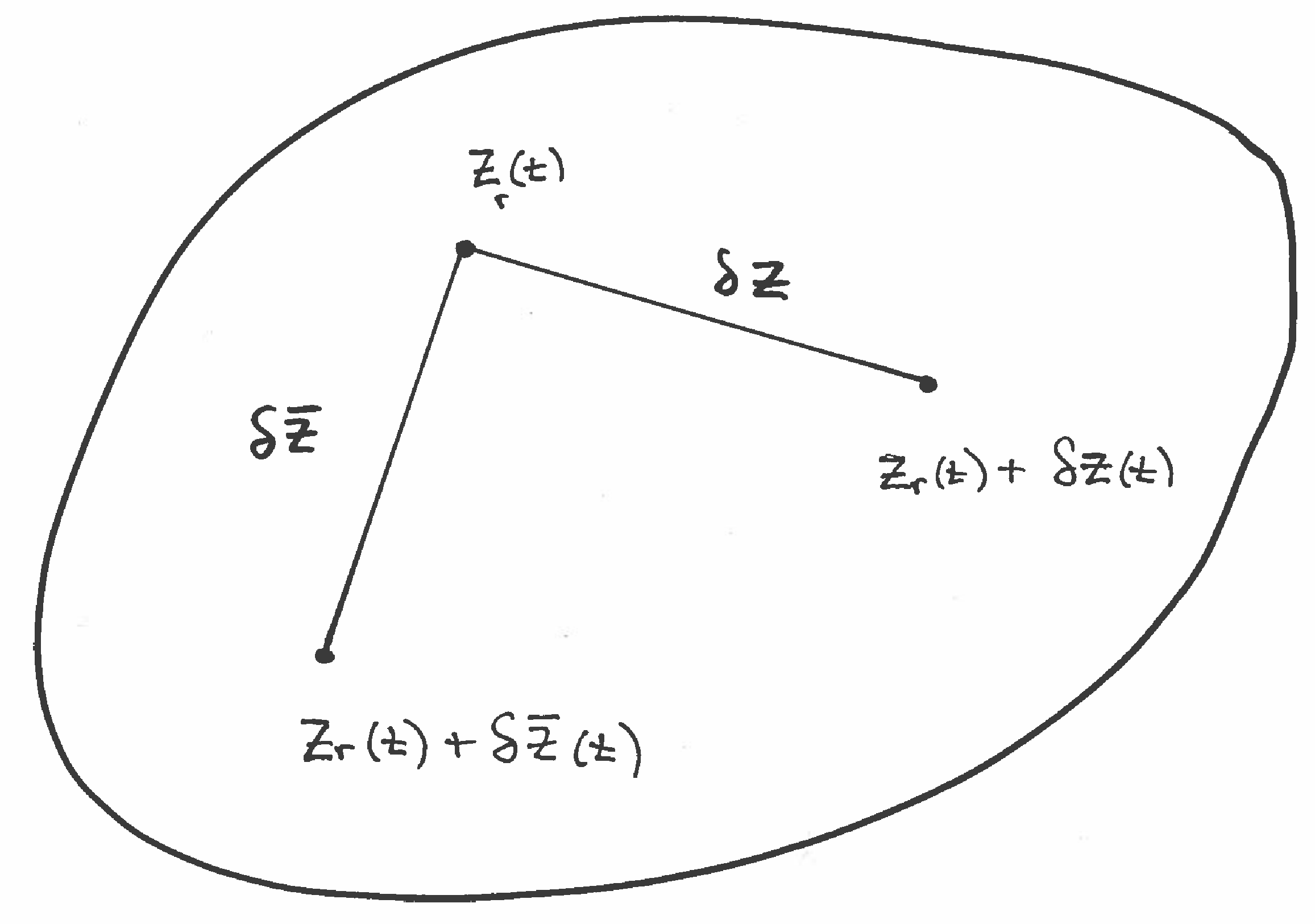}
\caption{Three nearby trajectories} 
 \label{fig:3traj}
\end{figure}

\begin{figure} 
  \centering
 \includegraphics[width=6cm]{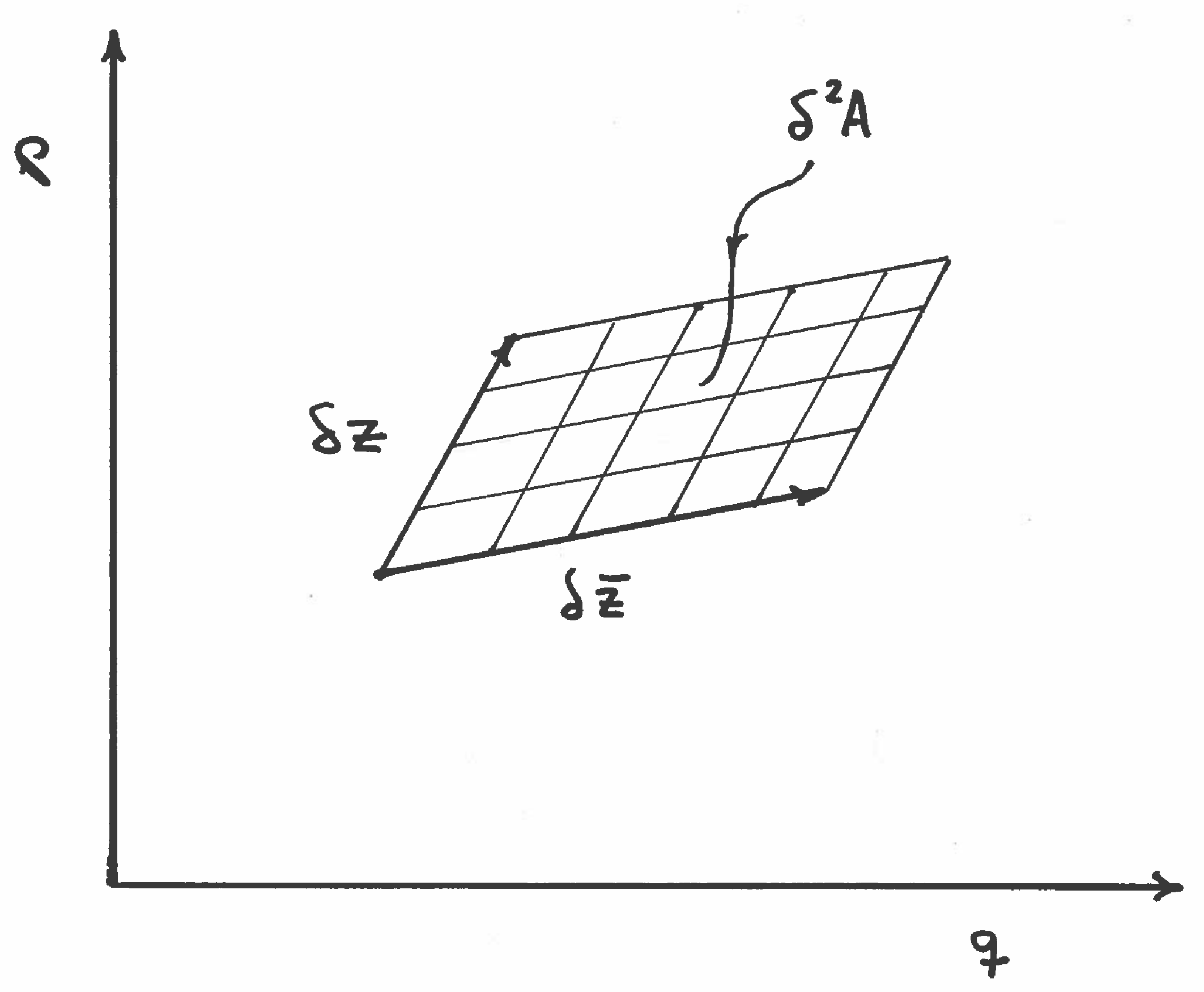}
\caption{Symplectic area} 
\label{fig:symA}
\end{figure}

To understand the importance of this main theorem, consider  three nearby trajectories, 
$z_r(t)$, $z_r(t)+ \de z(t)$, and $z_r(t)+ \de \bar{z}(t)$,  as depicted in Fig.~\ref{fig:3traj}
and let us investigate the following area-like quantity:
\bq
\de^2\!A=\de \bar{z}^{\al} \om_{\al \be}^c \de z^{\be}\,,  \quad \al,\be=1,2,\dots, 2N \,,
\nonumber
\eq
where
\bq
\om^c=
\left( 
\begin{array}{cc}
0_N & -I_N \\
I_N & 0_N  
\end{array}
 \right) = (J_c)^{-1}\,.
 \nonumber
\eq
A simple calculation for our three Hamiltonian trajectories reveals 
\bq
\frac{d}{dt} \, \de^2\!A=0 \,.
\label{sympA}
\eq
Thus, $\de^2\!A$, whatever it means,  is preserved by the Hamiltonian dynamics.  To understand its meaning, consider the case where  $N=1$.  Examination of  Fig.~\ref{fig:symA} reveals  that $\de^2A$ measures  the area of a parallelogram, i.e.,  $|\de^2 A|= |\de \bar{p}\de q- \de \bar{q}\de p|=|\de\bar{z}\times\de z|$. 
For $N> 1$ this quantity is a sum over such areas, $\de^2 A= \sum_i\left( \de \bar{p}_i\de q^i- \de \bar{q}^i\de p_i\right)$,  and is called the  first Poincar\'e invariant.  In more modern notation we would call this the symplectic two-form and write it  as $\om_c = \sum_i dp_i\wedge dq^i$.  In tensorial notation the canonical symplectic 2-form = (canonical Poisson matrix)$^{-1}$, i.e., $\om^c_{\al\ga}J_c^{\ga \be}=\de_{\al}^\be$.

There are consequences of \eqref{sympA}: upon summing (integrating) over the  infinitesimal areas $ \de^2\!A$ one sees that extended suitable two-dimensional ribbons in phase space are preserved under Hamiltonian dynamics.     Since an initial area in any $q-p$ subplane is preserved, one can form a $2N$-dimensional box, i.e., a volume that will be preserved. This is of course the famous Liouville theorem of Hamiltonian mechanics, which is a consequence of the preservation of  $\de^2 A$ but not vice versa.  In addition to regions of dimension 2 and $2N$, Hamiltonian  dynamics preserves everything in between, i.e.,  all the Poincare invariants which are surfaces of dimension 2, 4, 6, ... 2N.

Another family of invariants, intimately related to $\de^2 A$, are  the loop integrals
\bq
\calj=\oint_{\ga}p\cdot dq\,  \quad \mathrm{that \ satisfy}  \quad \frac{d}{dt} {\calj}=0\,, 
\label{loop}
\eq
where $\ga$ is  any closed curve  in phase space.  By a generalization of Stokes theorem,  these loop integrals and the  areas  $\de^2 A$ are related. Thus,  Hamiltonian dynamics has a  phase space circulation theorem, similar to that of  ordinary fluids.  

It should be borne in mind that if Hamiltonian dynamics did not have the properties described above, the nature of phase space would be quite different.  For example,  surfaces of section of magnetic field lines  would look very different, and lack the similarity we see.  Also, non-SI  of Hamiltonian systems can give rise to spurious damping or growth, a problem of concern when long time integration is desired. 

For infinite-dimensional canonical Hamiltonian  systems with conjugate fields $\chi=(\psi, \pi)$, \eqref{eqHam} becomes 
\bq
\psi_t=\frac{\de H}{\de \pi}= \{\psi, H\}
\quad \mathrm{and}\quad 
\pi_t=-\frac{\de H}{\de \psi}=\{\pi, \psi\}
\label{fieldHam}
\eq
where $H$ is now a Hamiltonian functional like the kinetic energy in a fluid $\int d^3x \rho\,  |\bfv|^2/2$,   partial derivatives are replaced by functional derivatives, and  the Poisson bracket operating on two functionals $F$ and $G$ is
\bq
\{F,G\}=\int\! d\mu\left(
\frac{\de F}{\de \psi} \frac{\de G}{\de \pi}-\frac{\de G}{\de \psi}\frac{\de F}{\de \pi}
\right)\,,
\label{canPB}
\eq
where the fields $\chi$ depend on $\mu$ as well as time.

\subsection{Symplectic Integrators}
\label{ssec:SI}
   Various approaches to symplectic integration have been followed.  Consider a  simple early example, starting from \eqref{Code} with a possibly time dependent vector field $V$.  Suppose one could find a time dependent coordinate change that takes this system into
 \bq
\dot{\bar{z}}=\bar{V}\equiv 0\,:
\label{bCode}
\eq
i.e., in the new coordinates the solution is constant in time,  $\bar{z}(\bar{z}_0,t)=\bar{z}_0$. Upon denoting such a time-dependent coordinate change by ${z}=\Phi(\bar{z},t)$, the initial conditions of the problem in the two system of coordinates are related by 
 ${z}_0=\Phi(\bar{z}_0,t_0)$, with $t_0$ being the initial time. If $z(z_0,t_0,t)$ were the solution to the original problem  \eqref{Code}, then the solutions to  \eqref{Code}  and \eqref{bCode} would in general be related by 
$\Phi\big(\bar{z}(\bar{z}_0,t_0,t),t\big)=z(z_0,t_0,t)$. 
But since the solution to   \eqref{bCode} is constant in time, we have 
\bq
z(z_0,t_0,t)=\Phi(\bar{z}_0,t)=\Phi\big(\Phi^{-1}({z}_0,t_0),t\big)\,,
\label{soltrans}
\eq
and the coordinate change is seen to solve  the problem.  

All of the above is moot unless one has a means of obtaining $\Phi$, but this can be attempted for Hamiltonian systems by making use of mixed variable generating functions, which not only provides an avenue for approximating $\Phi$ but does so in terms of a canonical transformation.  For clarity consider the case of a single degree of freedom and seek a canonical  transformation $z=(q,p)\leftrightarrow \bar{z}=(\bar{q},\bar{p})$.  This can be done by introducing a   mixed variable generating function of type three, $F_3(\bar{q},p,t)$, and giving the transformation as follows:
\bq
q=-\frac{\p F_3}{\p p}\qquad\mathrm{and} \qquad \bar{p}=-\frac{\p F_3}{\p \bar{q}}\,.
\label{mvgf}
\eq
Because the transformation has explicit time dependence, there is a  new Hamiltonian  given by
 \bq
\bar{H} = H +\frac{\p F_3}{\p t}
\label{newH}
\eq
and if this new Hamiltonian can be made constant or zero, then a Hamiltonian version of  \eqref{soltrans} applies.  Consider the simple case where $H=p^2/2 +V(q,t)$, with a time-dependent potential $V$, and choose $F_3= -\bar{q} \, p -t(p^2/2 +V(\bar{q},0))$.  By \eqref{newH} this gives rise to the new Hamiltonian
\bq
\bar{H}=V\big(\bar{q} + t(\bar{p} -t V_q(\bar{q},0)),t \big) - V(\bar{q},0)\,,
\label{snewH}
\eq
where $V_q=\p V/\p \bar{q}$.  Equation \eqref{snewH} gives upon expansion for small $t$, 
\bq
\bar{H}=t \big(
V_t(\bar{q},0) + \bar{p}  V_q(\bar{q},0)
\big) + O(t^2) \,,
\label{H0}
\eq
where $V_t =\p V/\p t$.  Because \eqref{H0} is $O(t)$ the same is true for Hamilton's differential equations, with the solutions
\bq
\bar{q} = \mathrm{const} +  O(t^2)
 \qquad \mathrm{and}\qquad 
\bar{p} = \mathrm{const} +  O(t^2) \,.
\nonumber
\eq
Thus, the map
\bq
\bar{p}= p+V_q(\bar{q},0) 
 \qquad \mathrm{and}\qquad 
 q = \bar{q} + pt
 \eq
is symplectic and  accurate to first order, which means that if we take $t$ to be the step size $\De t$, then the Taylor series in  $\De t$ of the approximate solution matches the actual to first order. 
 
 While the above example is rather simple, it serves to reveal a  path for obtaining higher order methods, such as the 3rd order method described in Ref.~\onlinecite{ruth83}  and  the higher order methods of  Refs.~\onlinecite{fr90,candy91,serna,feng,channell90,yoshida}. 
 
 One feature of transformations of given by \eqref{mvgf} is that these relations for general Hamiltonians are implicit and approximate solution  of  these formulas by e.g.\ Newton's method may lead to a violation of the symplectic property.  A similar difficulty arises with Lie transform approaches (see e.g.\ Ref.~\onlinecite{dragt82}), where the canonical transformation is generated by a series as follows:
 \bq
 \bar{z}=e^{\{G, \cdot\} }z= z + \{G, z\} +\frac12 \{G, \{G, z\} \} + \dots\,.
 \eq
 Truncation of this series again in general results in a transformation that  is not exactly canonical.   A way around this problem  is to make an approximation  using Cremona maps (see e.g.\  Ref.~\onlinecite{finn}). Using the notation $\{\cdot,g\}=X_g$ for a Hamiltonian vector field  one incorporates the Campbell-Baker-Hausdorff theorem to write 
\bq
 e^{\De t X_H}\approx\prod_i e^{\De t X_{H_i}}
 \label{lie}
 \eq 
where the $H_i$ are special polynomials for which the infinite series truncates, thus guaranteeing exact symplecticity.

 To summarize, every time step of a symplectic integrator is a CT and, consequently, the geometric structure described in Sec.~\ref{ssec:CHS} is exactly preserved.  However, it  is not the case that energy is exacly conserved as it is for CI.  Instead energy is `shadowed', which means  there is a nearby Hamiltonian that is exactly conserved.  In this way one can bound the deviations  of the energy, in accordance with the order of the scheme, and avoid energy drift (see e.g.\  Ref.~\onlinecite{HairerLubichWanner:2006}).  A  criticism of symplectic integration is that making  the time step adaptable can be difficult.  One  approach that overcomes this difficulty and allows for adaptive time steps is that of  Ref.~\onlinecite{richardson12}. 
 
For Hamiltonian field theories, PDEs that have canonical Poisson brackets of  the form of \eqref{canPB}, semidiscrete Hamiltonian reductions are  straightforward. One can simply expand the fields $(\psi,\pi)$ in terms of any complete set of basis functions and then truncate to obtain a finite-dimensional canonical Poisson bracket for ODEs, as is typical with spectral methods.  In this way it  is easy to show that the truncated system has  finite-dimensional Hamiltonian form. (See Refs.~\onlinecite{pjm81a,pjm81b} for an early  comparison of canonical  vs.\ noncanonical Hamiltonian reductions.)  Once the semidiscrete Hamiltonian  reduction is at hand, one can turn to SI as a possibility for  the time advancement of the ODEs.

  \section{Hamiltonian Integration (HI)}
  \label{sec:HI}
  
 Although SI as described in Sec.~\ref{sec:SI} is the most common type of integration that preserves Hamiltonian form, other methods and approaches exist.  In this section we discuss some of these.

 \subsection{Variational Integration (VI)}
 \label{ssec:VI}

The main idea behind VI  is to make approximations in an action functional, giving rise to time step equations as Euler-Lagrange equations.   In this way one can obtain symplectic integrators that may also preserve other invariants such as momenta.

 \subsubsection{Variational Structure - Canonical}
 
Consider the canonical case where we start  from  Hamilton's variational principle of mechanics. 
Recall,  Lagrange's equations arise from extremization of the following functional  over  paths:  
\bq
S[{q}] = \int_{t_0}^{t_1} L(q,\dot q,t)\, dt\,, \quad \delta q(t_0)=\delta q(t_1)=0\,,
\label{Saction}
\eq
where $L$ is the Lagrangian.  Thus, $\de S /\de q=0$ yields
\bq
\frac{\p L}{\p q^i} -\frac{d}{dt}\frac{\p L }{\p \dot q^i}=0\,.
\label{eqLag}
\eq
With the Lagrangian $L=\sum {m} |\dot q|^2/2 - \phi(q,t)$, Lagrange's equations are equivalent to Newton's second law. 

For convex Lagrangians, the  Legendre transform makes the identification of  Lagrange's $N$ second order equations with Hamiltonian's $2N$ first order equations.  Recall, one defines the canonical momentum by $p_i ={\partial L}/{\partial \dot{q}^i}$ and if  
\bq
W:=\det \left(\frac{\p^2 L}{\partial \dot{q}^i\p\dot{q}^j}\right)\neq 0\,.
\label{W}
\eq
then by the inverse function theorem one can solve for all the momenta $\dot{q}^i$  as a function of the $p_i$ and construct 
\bq
 H(q,p,t)= p_i \dot{q}^i  - L\,.
 \label{LegT}
 \eq
Then with the Hamilton of \eqref{LegT},  Hamilton's equations of  \eqref{eqHam} are equivalent to Lagrange's equations \eqref{eqLag}.  

When $W=0$ one cannot solve for all   the momenta $\dot{q}^i$  as a function of the $p_i$. For this more complicated situation  one must employ Dirac constraint theory (see e.g.\  Refs.~\onlinecite{sudarshan,pjmLB09}), which we will briefly consider in Sec.~\ref{sssec:VINC}.

 \subsubsection{Variational Integrators - Canonical}
 
For finite-dimensional systems, the  strategy for obtaining a variational integrator is to discretize the time integration of the action and  then vary.   As a simple example,  assume $t_1-t_0$ is small and use  the trapezoidal rule to approximate the action as 
\bqy
&&S[q]= \!\int_{t_0}^{t_1}\!\! L(q,\dot q,t)\, dt \approx\frac12(t_1-t_0)\bigg[
L\left(q_0, \frac{q_1-q_0}{t_1-t_0}, t_0\right) 
\nonumber\\
&&\hspace{.8 cm} + L\left(q_1, \frac{q_1-q_0}{t_1-t_0}, t_1\right)
\bigg]=W_d(t_0,t_1, q_0,q_1)
\eqy
where a  linear trajectory approximation has been used.  Next, we  add up to make a discrete action as follows: 
\bq
S_d = \sum_n W_d(t_n, t_{n+1}, q_n, q_{n+1})
\nonumber
\eq
Finally, upon variation, which now is merely differentiation, one obtains a  symplectic map for each time step.

There has been much work on VI  with an entry point into the literature being  Ref.~\onlinecite{MW01}. Historically, the idea to discretize actions is old and has been used for various purposes in Hamiltonian dynamical systems theory, both theoretically  and computationally. Theory work proposing VI  appeared in the 1980s (e.g.\ Ref.~\onlinecite{maeda80}), but  actual working codes for the more difficult problem of finding periodic orbits were given  in Refs.~\onlinecite{kook,Kook89,Meiss92}, where  use as an ODE integrator was also proposed.  

Variational integrators for PDEs proceed by discretizing in the spacial variable as well as time.  An example of this for MHD was given in Ref.~\onlinecite{zhou14} where discrete exterior calculus was used to  discretize the label of Newcomb's MHD Lagrangian.\cite{Newcomb62}

 \subsubsection{Variational Structure - Noncanonical}
 \label{sssec:VSNC}

Noncanonical variational  structure occurs when the Legendre transform fails. Consider the action functional
of the form of \eqref{Saction} with the Lagrangian
\bq
L= \dot{q}^a A_a(q) -\phi(q,t)\,,\qquad a=1,2,\dots, M
\eq
 which    would produce the canonical momenta $p_a=\p L/\p \dot{q}^a=A_a$.  For this case  one cannot solve for any of the  $\dot{q}^a$ in terms of the p$_a$ because the Lagrangian is totally degenerate, i.e., by \eqref{W}  $W\equiv 0$ with rank zero.  However, for this totally degenerate case the Euler-Lagrange equations are
 \bq
\om_{ba} \,\dot{q}^a= \frac{\p \phi}{\p q^b}\,,
 \eq
with $\om_{ba}:= A_{a,b} -   A_{b,a}$ where $A_{a,b}:={\p A_a}/{\p q^b}$. The  $A_a$ are  components of a 1-form and the 2-form $\om$ is closed,  $\om_{ba,c}+ \mathrm{cyc}\equiv 0$, where cyc means cyclic permutation  over $abc$.  If $\om$ has an inverse (which requires $M=2N$), say $J$, then the Euler-Lagrange equations are 
\bq
\dot{q}^a = J^{ab}(q)\,  \frac{\p \phi}{\p q_b}
\eq
and one has a `half-sized' Hamiltonian system in noncanonical variables with $\phi$ as the Hamiltonian. 

This example is of interest in plasma physics because  guiding center orbits are governed by 
  Littlejohn's Lagrangian\cite{littlejohn} (see also Ref.~\onlinecite{wimmel}) that has the following form: 
 \bq
L= \dot{\mathbf{x}} \cdot \big(\mathbf{A} + \mathbf{b}\,  v_{||}\big) - \frac12\,v_{||}^2 - \chi(\mathbf{x})\,.
\label{little}
\eq
where $\mathbf{A}(\bfx,t)$ is the  vector potential. 

 \subsubsection{Variational Integrators - Noncanonical}
 \label{sssec:VINC}

\begin{figure} 
\centering
\includegraphics[width=3.5in]{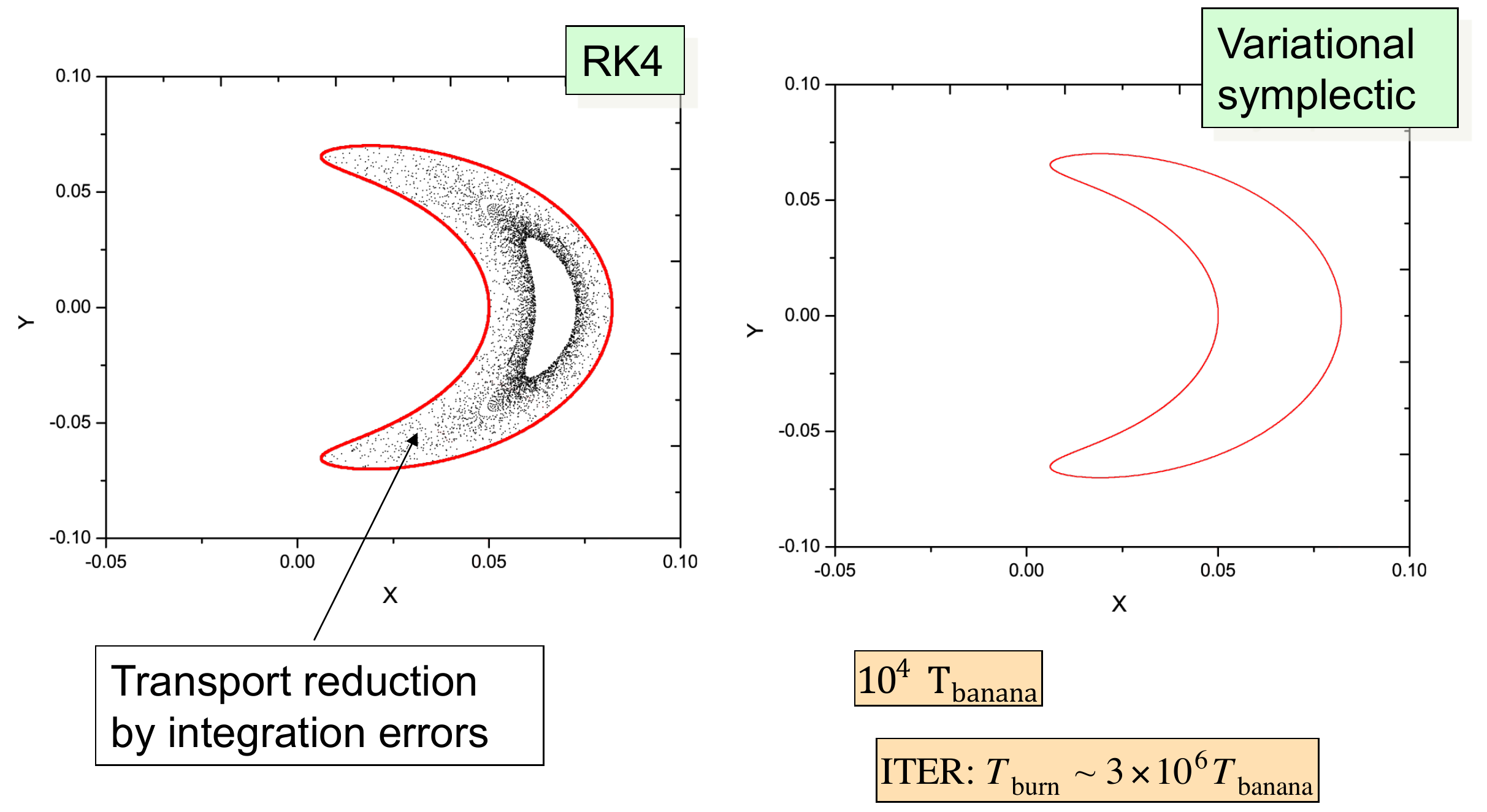}
\caption{Variational symplectic integrator vs.\ RKF for producing an accurate banana orbit. (Courtesy of H.\ Qin. See Ref.~\onlinecite{qin09}.) (left) Banana orbit using standard RK4 with exact orbit and (right) that obtained using a variational symplectic method. Orbits were obtained for ITER parameters with the integration time being $10^4$ banana periods.  Since the ITER burn time is more than  $10^6$ banana periods, numerical fidelity over very long times is required.} 
 \label{fig:HQ}
\end{figure}

One can use Dirac's constraint procedure\cite{dirac58} to obtain a canonical Hamiltonian system (see Ref.~\onlinecite{pjmP91} where this is done for \eqref{little}) but for the purpose of VI one can directly  discretize the action with \eqref{little}, as in Refs.~\onlinecite{qin08,qin09}, improved in Ref.~\onlinecite{ellison}, and  with ongoing work by J.\ Burby, J.\ Finn, M.\ Kraus, and H.\ Qin.   Figure \ref{fig:HQ} displays a guiding center orbit in tokamak geometry, with VI accuracy over a long time.

Plasma physics PDEs that model plasmas in terms of the usual  Eulerian variables are generally noncanonical in nature (see e.g.\ Ref.~\onlinecite{pjm05}).  Thus,  VI for such equations have subtleties.   In Ref.~\onlinecite{kraus13} this problem was handled for a variety of plasma models by introducing adjoint equations.  This procedure enables one to retain conservation laws on the discrete level.  Work on the Vlasov-Poisson system is recorded in Ref.~\onlinecite{KrausMajSonnendruecker:2016},  on MHD in Ref.~\onlinecite{tassi16}, and techniques for two-fluid theory also exist.\cite{pjmXQLYZH16}

  \subsection{Poisson Integration (PI)}
 \label{ssec:PI}

 In Sec.~\ref{sssec:VSNC} we observed that a Poisson matrix and consequently a Poisson bracket of nonstandard form (upon replacing of $J_c$ of \eqref{CPB} by $J$) may arise from a Lagrangian of noncanonical form.  Such unconventional Poisson  brackets have been explored since the nineteenth century, notably by Sophus Lie, and were further explored by many physicists and mathematicians in the twentieth century (e.g.\ Ref.~\onlinecite{sudarshan,souriau}).  In plasma physics, the Poisson bracket for motion of a particle in a given magnetic field was transformed to   noncanonical coordinates   for perturbation theory in Ref.~\onlinecite{littlejohn79}.  However, the word noncanonical was used in the seminal  Ref.~\onlinecite{pjmG80}  for a Poisson bracket that not only does not have the usual canonical form, but also possesses degeneracy.   Reference \onlinecite{pjmG80}  instigated  research on finite and infinite degree-of-freedom noncanonical Hamiltonian systems, particularly  plasma and other continuum models,  for which the map from Lagrangian to Eulerian variables results in a noncanonical degenerated Poisson bracket (see Ref.~\onlinecite{pjm98}).  The particular case of the Maxwell-Vlasov system  will be studied  in Sec.~\ref{sec:gempic}.

 \subsubsection{Noncanonical Hamiltonian Structure}
 
The noncanonical generalization of the Hamiltonian form of \eqref{zHam}  is given by 
\bq
\dot{z}^a=J^{ab}\frac{\p H}{\p z^b}=\{z^a,H\}\,,\qquad a=1,2,\dots, M
\label{NCham}
\eq
where the following noncanonical Poisson bracket   replaces \eqref{CPB}:
\bq
 \{f,g\}=\frac{\p f}{\p z^a}J^{ab}(z)\frac{\p g}{\p z^b}
 \,.
 \qquad a,b=1,2,\dots, M
 \label{GPB}
\eq
For a bracket of the form \eqref{GPB} to be a good Poisson bracket it must have the following properties  for all functions $f,g,h$:
\begin{itemize} 
\item  antisymmetry:
 \bq
 \{f,g\}=-\{g,f\} \quad \Leftrightarrow \quad J^{ab}=-J^{ba} 
 \label{antiS}
  \eq
\item  Jacobi  identity:
 \bq
\{f,\{g,h\}\}+  \ \mathrm{cyc}\equiv 0  
\quad \Leftrightarrow \quad J^{ad}J^{bc}_{,d} + \ \mathrm{cyc}\equiv 0\,,
\label{jacobi}
\eq
where cyc means cyclic permutation over $fgh$ in the first expression and over $abc$ in the second. 
\end{itemize}
This noncanonical generalization of the Hamiltonian mechanics is reasonable because of an old theorem due to 
Darboux, which states that if  $\det J \neq 0$ then there exists a coordinate  change   that (at least locally) brings $J$  into the canonical form $J_c$ of  \eqref{CPB}. Recalling the $J$ transform as a rank 2 contravariant tensor, this canonizing transformation $\bar{z}\leftrightarrow z$ would satisfy
 \bq
 {J}^{\mu\nu} \frac{\p \bar{z}^\al}{\p  {z}^{\mu}}\frac{\p \bar{z}^\be}{\p  {z}^{\nu}} = {J_c}^{\al\be}\,.
 \label{CJtrans}
 \eq
 However, the more interesting case is the one  studied by Sophus Lie where  $\det J= 0$.  This case  is degenerate and gives rise to Casimir invariants (Lie's distinguished functions), which are constants of motion for any possible Hamiltonian  that satisfy
 \bq
 \{f,C\}= 0 \ \forall \,f \quad \Leftrightarrow \quad J^{ab}\frac{\p C}{\p z^b}=0 \ \forall \,a\,.
 \eq
 Because of the degeneracy, there is no coordinate transformation to canonical form; however,  a theorem known to Lie (see e.g. Refs.~\onlinecite{eisenhart,littlejohn82})  which we call the Lie-Darboux theorem states  there is a transformation to the following degenerate canonical form:
 \begin{equation}
 J_{dc}= \left(
 \begin{array}{ccc} \,0_N&\,I_N&\,\,\, 0\\
 -I_N&\,0_N&\,\,\, 0\\
 0&\,0&\,\,\, 0_{M-2N}\\
 \end{array} \right)\,.
 \label{III-9-3}
 \end{equation}
 
\begin{figure}[htb]
\includegraphics[height=1.8 in,width= 2.7in]{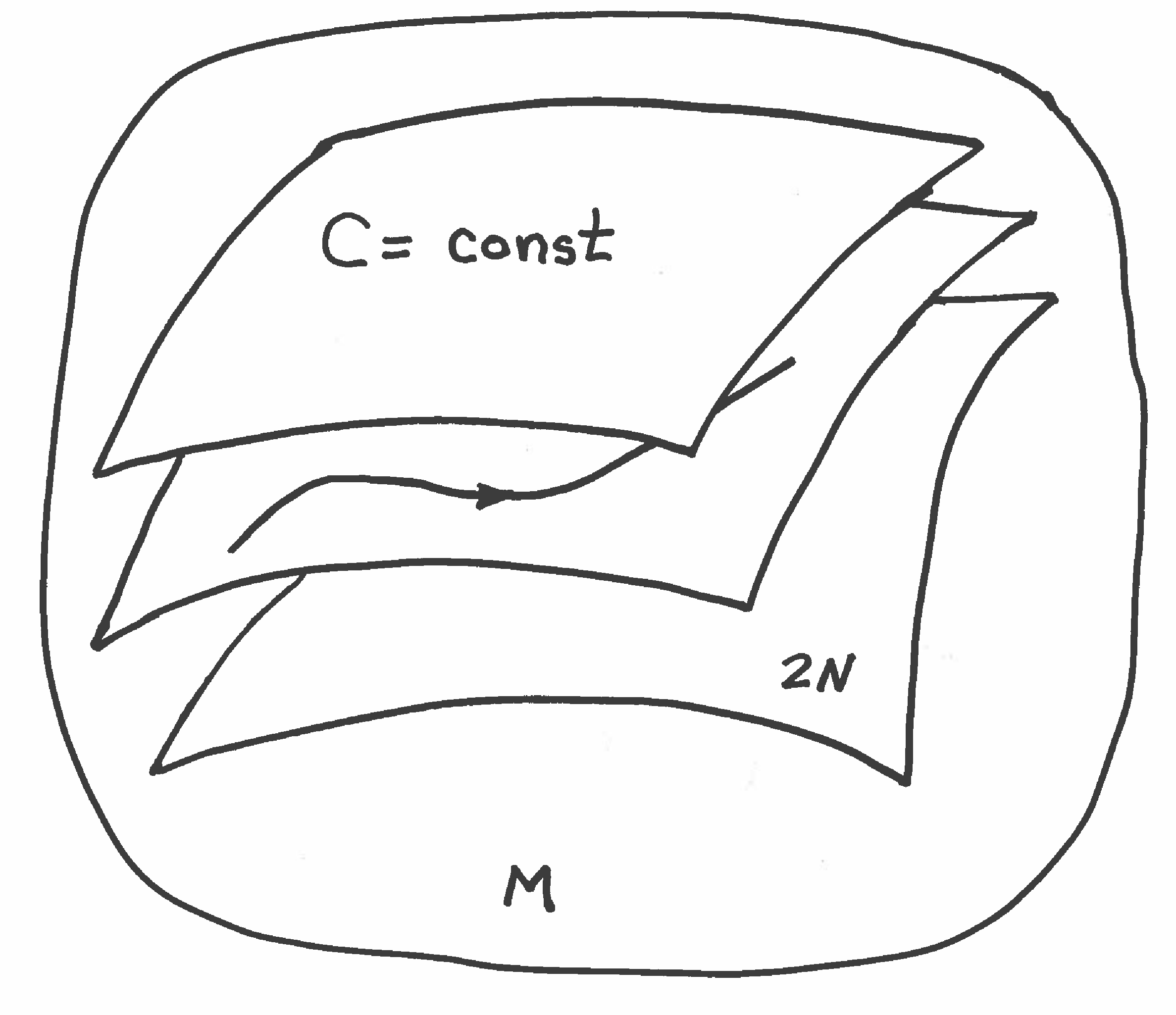} 
\caption{Depiction of a Poisson manifold foliated by symplectic leaves of constant Casimir invariants.}
\label{fig:poissman}
\end{figure}

Instigated in a major way by the noncanonical Poisson brackets for plasma models, manifolds with the addition of degenerate Poisson bracket structure, known as Poisson manifolds,  have now been widely studied (see e.g.\ Ref.~\onlinecite{weinstein83}).   The local structure of a Poisson manifold is depicted in Fig.~\ref{fig:poissman},  where it is seen that the phase space is 
foliation by  the level sets of the Casimir invariants.  For an $M$-dimensional system there exist $M-2N$ Casimir invariants, and an orbit that initially lies on such a surface defined by the level sets of the initial Casimir invariants remains there.  These surfaces, called symplectic leaves, have dimension $2N$ and  the phase space is generically foliated by them. 
 \medskip

Lie-Poisson brackets are a special form of noncanonical Poisson bracket that typically appear in matter models in terms of an Eulerian variable description.   For finite-dimensional Lie-Poisson Hamiltonian systems,  the Poisson matrix  $J$ is linear in the dynamical variable and  has the  form $J^{ab}= c^{ab}_cz^c$,  where the numbers $c^{ab}_c$ are the structure constants of some Lie algebra. 

Noncanonical Hamiltonian field theories have Poisson brackets of the form
\bq
\{F,G\}=\int\! d\mu \, 
\frac{\de F}{\de \chi^i} \mathcal{J}^{ij}\frac{\de G}{\de \chi^j} 
\,,
\label{NcanPB}
\eq
where $F$ and $G$ are arbitrary functionals,   the Poisson matrix becomes a Poisson operator  $\mathcal{J}$, and the   fields $\chi=(\chi^1, \dots, \chi^N)$ are not divided into canonical pairs $(\psi,\pi)$ as in \eqref{canPB}.   In general the Poisson operator  may depend on the fields $\chi(\mu,t)$,  but most importantly $\mathcal{J}$ must still satisfy infinite-dimensional versions of \eqref{antiS} and \eqref{jacobi}, the latter of which may be challenging to show  (see Ref.~\onlinecite{pjm82}).

A example of an  infinite-dimensional noncanonical Lie-Poisson bracket is that for the Vlasov-Poisson system, \cite{pjm80,pjm82,pjm81b} 
  \bq
  \{F,G\}=\int\! dxdv\,  f \left[ \frac{\de F}{\de f}, \frac{\de G}{\de f}\right]
  =\int\! dxdv\,  \frac{\de F}{\de f}\mathcal{J}\frac{\de G}{\de f}\,,
  \label{VPbkt}
  \eq
  where  $f(x,v,t)$ is the distribution function  and the Poisson operator for  this case is given by 
  \[
  \mathcal{J}\, \cdot =- [f,\cdot\,]=f_\bfx\cdot\p_\bfv \cdot -f_\bfv\cdot \p_\bfx \, \cdot
  \]
Here the value of the electron charge and mass have been set to unity and consequently the Hamiltonian  of the system, the total energy, is given by
    \bq
    H[f]= \int\! dx dv \,f {|\bfv|^2}/2+ \int\! d{ x}\, |\nabla \phi|^2/2
    \label{VPham}
    \eq
    where$ |\nabla \phi|^2$ is a shorthand for the quadratic Green's function expression in terms of $f$.  Using   $\de H/\de f=v^2/2 -\phi(f;x)=\mathcal{E}$,  the  particle energy, the Vlasov-Poisson system is expressed in Hamiltonian  form as
   \[
   f_t=\{f,H\}=-[\cale, f]\,.
   \]
 In Sec.~\ref{sec:gempic} we will see that the Poisson bracket of \eqref{VPbkt}  is a part of the more complicated  full Vlasov-Maxwell Hamiltonian structure.   
 
 \subsubsection{Poisson Integrators}
 
 Symplectic integrators preserve, step-by-step,  the canonical Poisson matrix $J_c$ (cf.~\eqref{Jtrans}).  Thus, it would seem natural  to  consider transformations $\bar{z} \leftrightarrow z$ that preserve the form of a given noncanonical Poisson matrix, i.e. transformations for which
 \bq
{J}^{\mu\nu}({z}) \frac{\p \bar{z}^\al}{\p {z}^{\mu}}\frac{\p  \bar{z}^\be}{\p  {z}^{\nu}}=  {J}^{\al\be}(\bar{z})\,.
\label{JncTrans}
 \eq
Such transformations are called Poisson maps.   If an integrator is constructed to maintain \eqref{JncTrans}  step-by-step, then up to coordinate changes there is a preserved canonical $J_c$ and one has all the structure of SI maintained. To see this, suppose \eqref{JncTrans} is written symbolically as $\tilde{T} J T= J$, where $T=\p \bar{z}/\p z$ and tilde denotes transpose.  Because of the Lie-Darboux theorem we know there is another coordinate  change, say $M$,  that `canonizes' $J$, i.e., $\tilde{M} J M= J_{dc}$.   A straightforward calculation shows that the transformation $N:=M^{-1} T M$ is symplectic up to Casimirs, $\tilde{N} J_{dc} N= J_{dc}$.
 Thus, deep down there is a preserved 2-form and on a symplectic leaf one has SI.   In addition, if one advances time with a $T$ that satisfies \eqref{JncTrans}, then because the Poisson manifold geometry depicted in Fig.~\ref{fig:poissman} is coordinate independent,  the Casimir invariants will all exactly be preserved. 

Examples of works  on PI for  Lie-Poisson ODE systems are Refs.~\onlinecite{channell91,rob,engo}  and the associated development of Lie group  integrators was considered in Ref.~\onlinecite{celledoni}.

For noncanonical Hamiltonian field theories, PDEs that have canonical Poisson brackets of  the form of \eqref{NcanPB} with $\mathcal{J}$ depending on $\chi$, semidiscrete Hamiltonian reductions are not straightforward.  In fact, 
prior to the publication of Ref.~\onlinecite{pjmG80} considerable effort was spent on trying to find a finite-dimensional Hamiltonian projection of MHD by expansion in a  complete set of basis functions followed by various truncations,  but it was learned that such a procedure results in the  failure of the Jacobi identity.  This is more easily seen for the simpler case of \eqref{VPbkt}, which was tried in Ref.~\onlinecite{pjm81a} by expansion in  a Fourier basis --  there  it was noted for this projection that  the process of truncation destroys the Jacobi identity.  However, it was also noted there that expansion in terms of particle degrees of freedom results in a good Poisson bracket.    This procedure will be used in Sec.~\ref{sec:gempic} where we will discuss GEMPIC, which is a kind of  PI in the PDE context for the full  Maxwell-Vlasov system.

  \section{Metriplectic Integration (MI)} 
  \label{sec:MI}
  
  As noted in Sec.~\ref{sec:CI} reductions of the general HAP formalism  of \eqref{kinetic}--\eqref{field} and its associated Hamiltonian description may  give rise to dissipation, which generally results  in  vector fields that have Hamiltonian and non-Hamiltonian or dissipative  components.  Generally speaking, dissipation takes one of two forms in continuum matter models that describe media,  such as fluids and plasmas.  The first form is exemplified by the Navier-Stokes equation where viscosity removes energy from the system, while the second form  is exemplified by the  fluid theory that conserves energy but allows for viscous heating, thermal diffusion, and entropy production by using a general equation of heat transfer.\cite{landau}  Transport equations with collision operators,   such as  the  Boltzmann, Landau-Fokker-Planck, or gyrokinetic\cite{brizardAPS} collision operators,  are of the second form because they conserve mass,  momentum, and energy, but produce entropy.  Metriplectic dynamics\cite{pjm84,pjm86} is general formalism that embodies systems like these that in a real dynamical sense embody both the first and second laws of thermodynamics, i.e., have  energy conservation and entropy production.

  \subsection{Metriplectic Structure}
  \label{ssec:MS}
  
The theory of metriplectic dynamics  evolved out of  early considerations of combining dynamics generated by Poisson brackets  together with dissipative effects,\cite{pjmK82,pjmH84}  the near simultaneous publications of Refs.~\onlinecite{kaufman84,pjm84,pjm84b,grmela84},  with the complete set of axioms as described in this section  first appearing in Refs.~\onlinecite{pjm84,pjm84b}.  The name metriplectic dynamics for this set of axioms was introduced in Ref.~\onlinecite{pjm86} (and renamed GENERIC in Ref.~\onlinecite{grmela97}).  There has been much subsequent work, notably by Grmela and collaborators,\cite{ottinger} and myself and others.\cite{pjm09a,pjmBR13,tassiM,materassi,guha}

A metriplectic vector field has two parts, a part that is  Hamiltonian and a part that is a degenerate gradient or metric flow.   Consider first a metric  flow on a finite-dimensional phase space manifold, which in coordinates has the form
\bq
\dot{z}^a=g^{ab}\frac{\p S}{\p z^b}=(z^a,S)\,,\qquad a=1,2,\dots, M
\label{Mflow}
\eq
where $S$ is an `entropy' function and  the  symmetric  bracket  $(f,g)=(g,f)$,  defined on arbitrary phase space functions $f$ and $g$,  is defined by 
\bq
 (f,g)=\frac{\p f}{\p z^a}g^{ab}(z)\frac{\p g}{\p z^b}
 \,.\qquad a,b=1,2,\dots, M
 \label{SPB}
\eq
Metriplectic dynamics requires two things of the matrix $g$:  (i) that it be positive semi-definite so that the dynamics satisfies
 \[ 
 \frac{d S}{dt}   =(S,S)\geq 0\,,
 \]
 which can be used to build-in asymptotic stability, i.e.  an  `$H$-theorem', and (ii) degeneracy so as to conserve an energy $H$ that will act as a Hamiltonian, 
  \[
(H,f)=0 \  \ \forall f\,.
\]  

Paired with the gradient flow is a noncanonical Hamiltonian flow of the form of  \eqref{NCham} with Poisson bracket \eqref{GPB}, 
 and the two together define a  metriplectic vector field  generated by some function $\mathcal{F}$ as follows:
 \bq
 \dot{z}^a= J^{ab}\frac{\p \mathcal{F}}{\p z^b}  + g^{ab}\frac{\p \mathcal{F}}{\p z^b}\,.
 \eq
 The  metric and Hamiltonian components are then mated by requiring $\mathcal{F}=H+S $ with the entropy $S$ selected from the set of Casimir invariants of the noncanonical Poisson bracket.  Given this structure we have the following:
\begin{itemize}
\item a 1st Law: 
\[
\dot H=\{H,\calf\}+ (H,\calf)= 0+ (H,H) + (H,S)= 0
\]
where we  identify energy with the Hamiltonian $H$  

\item a 2nd Law: 
\[
\dot S= \{S,\calf\}+ (S,\calf)= (S,S)\geq 0
\]
where we  identify entropy  with a  Casimir $S$, and  entropy production yields  Lyapunov relaxation to equilibrium; i.e., the dynamics effects  the variational principle, $\de \calf=0$.  Observe  $\de \calf=0$ is the energy-Casimir variational principle (see e.g.\ Ref.~\onlinecite{pjm98}) and the choice of `thermal equilibrium' is determined by the choice of Casimir invariant as entropy. 
\end{itemize}
Thus metriplectic dynamics is a    dynamical paradigm  that embodies both the first and second laws of thermodynamics. 
 
A finite-dimensional example of a metriplectic system based on the Hamiltonian structure of the free rigid body was given in Ref.~\onlinecite{pjm86}.   For this example, the energy of the rigid body is conserved, but the magnitude of the angular momentum, which acts as the entropy,  monotonically changes until the system relaxes to one of its stable rotations about a principal axis. 

A general infinite-dimensional  symmetric bracket has the  form
\bq
(F,G)=\!\int\! d\mu' \! \!\int\! d\mu''\, 
\frac{\de F[\chi]}{\de \chi^i(\mu')} \mathcal{G}^{ij}(\mu',\mu'')\frac{\de G[\chi]}{\de \chi^j(\mu'')}\,, 
\label{gensymbkt}
\eq
where for metriplectic dynamics $\mathcal{G}^{ij}$ is chosen to guarantee positive semidefiniteness $(F,F)\geq0$, the symmetry $(F,G)=(G,F)$, and   to  build-in the degeneracy condition $(H,F)=0$ for all functionals $F$.  
 
A most  important class of metriplectic systems are transport equations of the form 
 \bq
 \frac{\p f}{\p t}= -\bfv\cdot\nabla f + \bfE\cdot \nabla_v f +\left. \frac{\p f}{\p t}\right|_c,
 \label{genCol}
 \eq 
where the first two terms on the right-hand side of \eqref{genCol}, the Hamiltonian Vlasov terms for electrons, can be generated by  the Poisson bracket of \eqref{VPbkt}, while the metriplectic formalism is completed by using the following to generate  the  collision operator,   ${\p f}/{\p t}|_c$:\cite{pjm86}
\begin{eqnarray}
(F,G)&=&\int\!dz\int\!dz'\left[
\frac{\p}{\p v_i}\, \frac{\delta F}{\delta f(z)}
- 
\frac{\p}{\p v'_i}\, \frac{\delta F}{\delta f(z')}
\right]
\label{metriBkt}\\
&& \hspace{.5 cm} 
 \times\,  T_{ij}(z,z')\left[
\frac{\p}{\p v_j}\, \frac{\delta G}{\delta f(z)}
- 
\frac{\p}{\p v'_j}\, \frac{\delta G}{\delta f(z')}
\right]\,.
\nonumber
\end{eqnarray}
In \eqref{metriBkt} $z=(\bfx,\bfv)$, $i,j=1,2,3$,  and it remains to tailor this expression to fit  a specific collision operator.   Because Casimir invariants are candidate entropies, 
\[
S[f]=\int\!dz \, s(f)
\]
and the specific choice of the entropy density function $s$ will determine the state to which the system relaxes, i.e., the state of thermal equilibrium.  If the tensor $T_{ij}$ is given by 
\[
T_{ij}(z,z')=w_{ij}(z,z') M(f(z))M(f(z'))/2
\]
with the functions $M$ and $s$ related by the following entropy compatibility condition:
 \[
  M \frac{d^2s}{d f^2}= 1\,,
\]
then it only remains to determine the tensor $w_{ij}(z,z')$.  Any choice that satisfies the symmetry conditions 
\bqy
&& w_{ij}(z,z')=w_{ji}(z,z')\,,  
\qquad   w_{ij}(z,z')=w_{ij}(z',z)\,,
\nonumber\\
 && g_iw_{ij} =0\,,
 \qquad \mathrm{with}\qquad   g_i= v_i-v_i'\,,
\eqy
will assure conservation of mass, momentum, and energy.   Finally, according to the metriplectic prescription, the system will relax to the state determined by extremization of $\mathcal{F}=H + C$, i.e., 
\bq
\mathcal{E} +\frac{ds}{df}= 0
\label{Mequil}
\eq
where recall $\mathcal{E}= \de H/\de f$.  If the solution to \eqref{Mequil} is stable, which is assured by the Kruskal-Oberman\cite{KO} monotonicity stability condition, then at least on a formal level $dS/dt\geq 0$ will cause $S$ to increase until the monotonic state $f=(ds/df)^{-1}(-\mathcal{E})$  is achieved.   Thus, this collision operator can be tailored so that the system relaxes to {\it any}  distribution function that is monotonic in the energy.   Proving  the  relaxation property is detailed,  but it can be shown  by paralleling the arguments of Ref.~\onlinecite{lenard}.
 
The Landau collision operator is obtained by choosing the kernel
\[
w_{ij}^{(L)}=(\delta_{ij}-g_ig_j/g^2) \delta(\mathbf{x}-\mathbf{x}')/g\,,
\]
the entropy density $s=-kf\ln f$, and  the entropy compatibility condition gives 
  \[
 M \frac{d^2s}{d f^2}= 1 \Rightarrow M\propto f\,.
\]
Another choice is that of Kadomstev and Pogutse of Ref.~\onlinecite{KP70} who attempted to explain relaxation to a 
Lynden-Bell (or Fermi-Dirac) equilibrium state. For their collision operator  $s=-k[f\ln f + (1-f)\ln(1-f)]$ and  the entropy compatibility condition yields 
  \[
  M \frac{d^2s}{d f^2}= 1\quad  \Rightarrow \quad  M\propto f(1-f)\,.
\]
In this way, one can construct a collision operator that relaxes to any  equilibrium state monotonic in the energy.

 \subsection{Metriplectic Integrators}
 \label{ssec:MI}
 
 MI should at once have a  PI  component for the Hamiltonian part of its vector field, preserving the symplectic nature of phase space and the Casimir foliation,  while it should have a CI component for the dissipative part that exactly conserves invariants.  
 Various ideas come to mind on how to achieve this, e.g.\  by splitting the time step.  At present we know of no published metriplectic integrators, finite or infinite.  However, very recently progress has been made and  examples of MI are on the horizon.\cite{MKEH, KKMS16c}

 \section{Simulated Annealing (SA)}
 \label{sec:SA}
 
By simulated annealing we mean a numerical relaxation procedure where structure is used to obtain physical equilibrium states by constructing usually nonphysical dynamics that possess certain geometrical constraints.  In this section we discuss two kinds of SA:  one  based on the metriplectic dynamics discussed in Sec.~\ref{sec:MI} and the other called double bracket dynamics (introduced in Ref.~\onlinecite{vallis89}), which  is also based on Hamiltonian structure.

  \subsection{Symmetric Bracket Structure and  SA}
  
 Both metriplectic and double bracket  SA use a symmetric bracket to achieve  relaxation.  Such brackets can be constructed in a variety of ways, giving rise to a bracket of the form of \eqref{gensymbkt} or its finite-dimensional analog.  Since such brackets are interesting independent of their SA application, we explore their construction in some detail. 
 
 One way of building in degeneracy is with brackets  of   the following form:  
 \bq
(F,G)= \!\int\! d\mu' \! \!\int\! d\mu''    \{F,\chi^i(\mu')\} \calk_{ij}(\mu',\mu'') \{\chi^j(\mu''),G\} \,,
\label{Gsymbkt}
\eq 
where   $\mathcal{K}^{ij}$ is positive semidefinite  and $\{\,,\,\}$  is any Poisson bracket.  This form,  given in Ref.~\onlinecite{pjmF11},  has a very general geometric significance.\cite{pjmBR13}   To see this suppose $P$ is  any phase space manifold that has both Hamiltonian and Riemannian structure.  Because  $P$ has Riemannian  structure,   given any  two vector fields $X_{1,2}\in\mathfrak{X}(P)$  the following is defined: 
$$
\mathbf{g}(X_1,X_2):\mathfrak{X}(P)\times \mathfrak{X}(P)\rightarrow \R\,.
$$
If the two vector fields are Hamiltonian,  in terms of either a degenerate or a nondegenerate Poisson matrix, say $X_{F},X_{G}$, then we natually have a symmetric bracket given by 
\bq
(F,G)=\mathbf{g}(X_F,X_G)\,,
\label{Gnat}
\eq
which is an abstract way of writing \eqref{Gsymbkt}.  Thus, natural symmetric brackets exist for K\"ahler manifolds, which naturally have  metriplectic or double bracket flows. 

If  $P$  is a  manifold with any degenerate Poisson bracket, then the associated Casimir invariants $C$ must satisfy   $(F,C)\equiv 0$ for all $F$.  One possibility is the case where  $\{\, ,\, \}$ is  a Lie-Poisson bracket.  Another possibility is to build a Dirac bracket with desired degeneracies,   as was done in Ref.~\onlinecite{pjmF11} for SA.   Given any bracket, canonical or not, one can build in Casimir invariants by Dirac's construction.  This is easily  done  in finite dimensions for the case of building in  invariants, say $D_{i}$.  For this case  one calculates the matrix  $\Om_{ij}=\{D_i,D_j\}$,  which must be invertible, whence the Dirac bracket is given by 
\bq
\{F,G\}_*=\{F,G\} - \{F,D_i\} (\Om^{-1})^{ij}  \{G,D_i\}\,,
\eq
where it is easily seen that $\{F,D_i\}_*= 0$ for all $F$. (See  Refs.~\onlinecite{pjmLB09,pjmCT12,pjmCGBT13,pjmCT14} for discussion and use  of Dirac brackets in the  infinite-dimensional context.)  A third possibility is to use the triple bracket construction introduced in Refs.~\onlinecite{pjmM84,pjmB91} and considered again  in Ref.~\onlinecite{pjmCGBT13}.  For example, suppose $\{A,B,C\}$ is a  purely antisymmetric triple bracket  on functionals $A,B,C$. Then, if one desires a symmetric bracket that conserves  a desired Hamiltonian, one can use $\{F,G\}_H:=\{F,G,H\}$ in \eqref{Gsymbkt}.  Similarly, one could use antisymmetric brackets with more slots.  A triple bracket will be used for metriplictic SA  in Sec.~\ref{sssec:MSA}.

Metriplictic SA proceeds just as in Sec.~\ref{sec:MI}, by employing a metriplectic dynamical system to numerically effect  the variational principle $\de \calf =\de (H + S)=0$, and  finds  the state consistent with the initial energy or other invariants.  On the other hand, double bracket SA extremizes $H$ subject to constancy of all Casimir invariants.  We will give example of these in Secs.~\ref{sssec:MSA} and \ref{sssec:DBSA}, respectively.

\subsection{Examples of Simulated Annealing}

\subsubsection{Metriplectic SA}
\label{sssec:MSA}

As a numerical tool, metriplectic SA is in its infancy.   The work described in this section is joint work with Glenn Flierl\cite{pjmFl15,pjmFl16}  and is preliminary in nature.  Additional preliminary yet promising results have been obtained with and by collaborators at IPP in Garching\cite{pjmBKMS16} where this approach is being explored for obtaining MHD equilibria.  It is important to clarify that metriplectic SA differs  from the MI of Sec.~\ref{ssec:MI} in purpose:  metriplectic SA  is a tool for finding equilibrium states via, in general, unphysical dynamics, while MI is an integrator that preserves the MI geometric structure. 

The example at hand concerns quasigeostrophic flow with topography, as governed by the following PDE for the potential vorticity $\ze(x,y,t)$:
\bq
\ze_t=[\ze,\psi]
\eq
where in the present section $[f,g]=f_xg_y-f_yg_x$ and the stream function $\psi$ satisfies  $\ze=\nabla^2 \psi +T$, with $T(\mathbf{x})$ modeling bottom  topography.  (See e.g.\ Ref.~\onlinecite{pedlosky}.)  Observe, this system also has an interpretation  for guiding center plasmas and is closely related to the Hasegawa-Mima equation for plasma drift waves.\cite{horton}

Our goal is to find an asymptotic time-independent state consistent with conservation of energy.  We seek this state by a procedure that extremizes the enstrophy, in particular,  we will construct a metriplectic system that minimizes 
\bq
Z=\int d^2x\, \ze^2/2\,.
\eq
(Note, with a flip of a sign it can be made to maximize $Z$, which is not important since  a physical process is not being modeled.)
  
We construct our symmetric bracket by means of the following antisymmetric triple bracket:\cite{pjmM84,pjmB91}
\bq
\{A,B,C\}=\int \!d^2x\,  A_\ze[B_\ze,C_\ze] 
\label{qgtriple}
\eq
where $A,B,C$ are arbitrary functionals and we use the notation $A_\ze:=\de A/\de \ze$.  That \eqref{qgtriple} is completely antisymmetric is easily shown for the periodic boundary conditions we use by integration by parts. 
From \eqref{qgtriple}  we construct the symmetric bracket, 
 \bqy
 (F,G)&=&  \int\!d^2x\int\!d^2x'\, \{\ze(x,y), F, H\}\,  
\label{symQG}\\
 &&\times \ \mathcal{K}(x,y|x',y') \,  \{\ze(x',y'), G, H\}  \,,
\nonumber
 \eqy
which clearly satisfies $(F,H)=0$ for all $F$. From \eqref{qgtriple}, \eqref{symQG} reduces to 
 \bq
  (F,G)= \int\!d^2x\int\!d^2x'\, [F_\ze, H_\ze] \calk(x,y|x',y') [G_{\ze'}, H_{\ze'}]' \,. 
\nonumber
  \eq

Now we choose $H=-\int d^2x \,\psi \ze/2$, the physical quasigeostrophic energy, and suppose 
\bq
\calk(x,y|x',y')=-\ga \de(x-x')\de(y-y') 
\eq
giving rise to 
\bq
\frac{\p \ze}{\p t} = (\ze, Z)= 
\ga\big[\ps,[\ps,\ze]\big]\,, 
\nonumber
\eq
which coincidentally is an equation of the form of that in Ref.~\onlinecite{sadourny}, which was argued to be of physical origin. 

\begin{figure}[htb]
\centering
\subfigure[{\footnotesize \  $t=0$.}]{\includegraphics[width=0.23\textwidth]{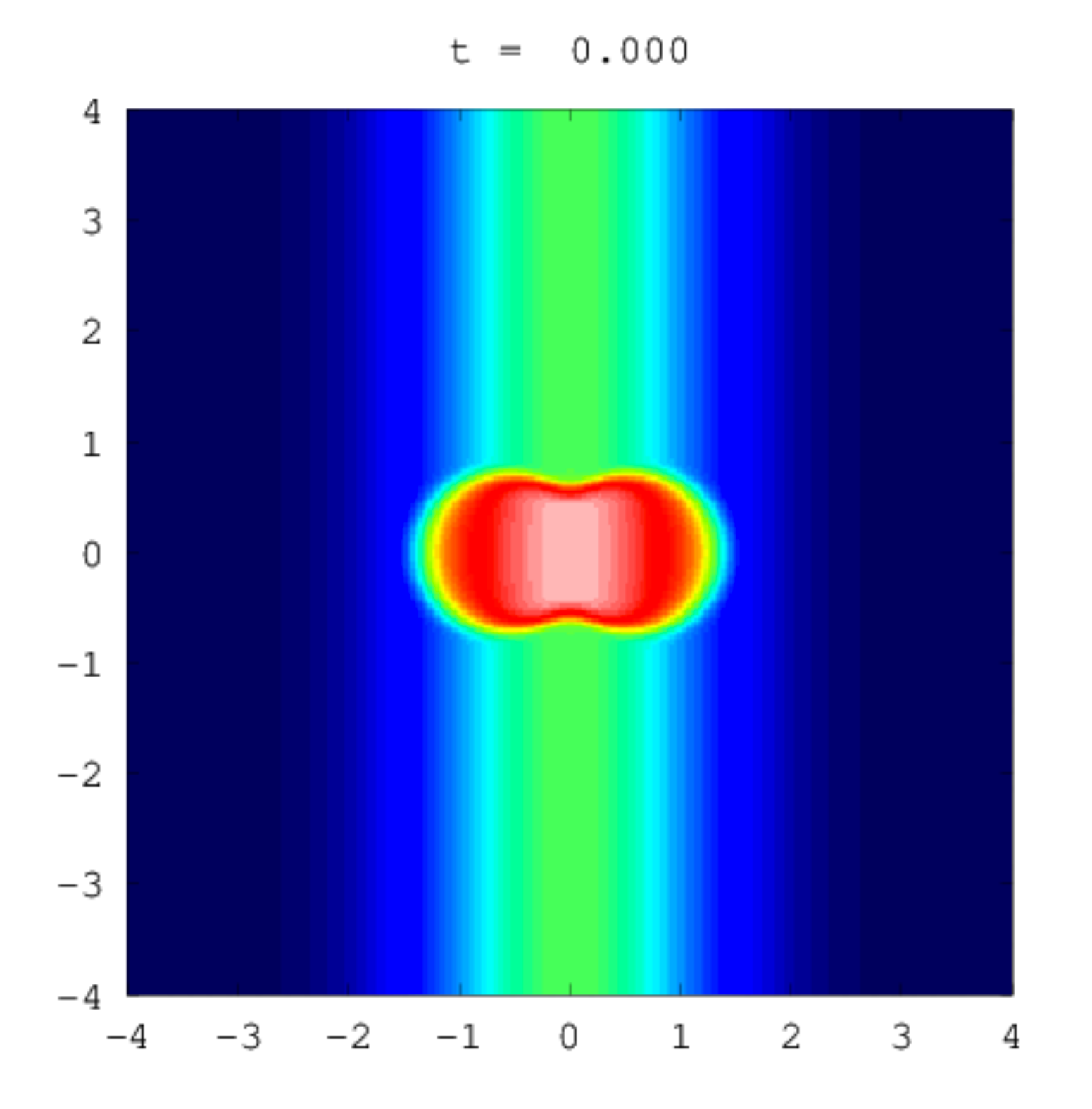}
\label{fig:MSA0}
}
\subfigure[{\footnotesize \  $t=10$.}]{\includegraphics[width=0.23\textwidth]{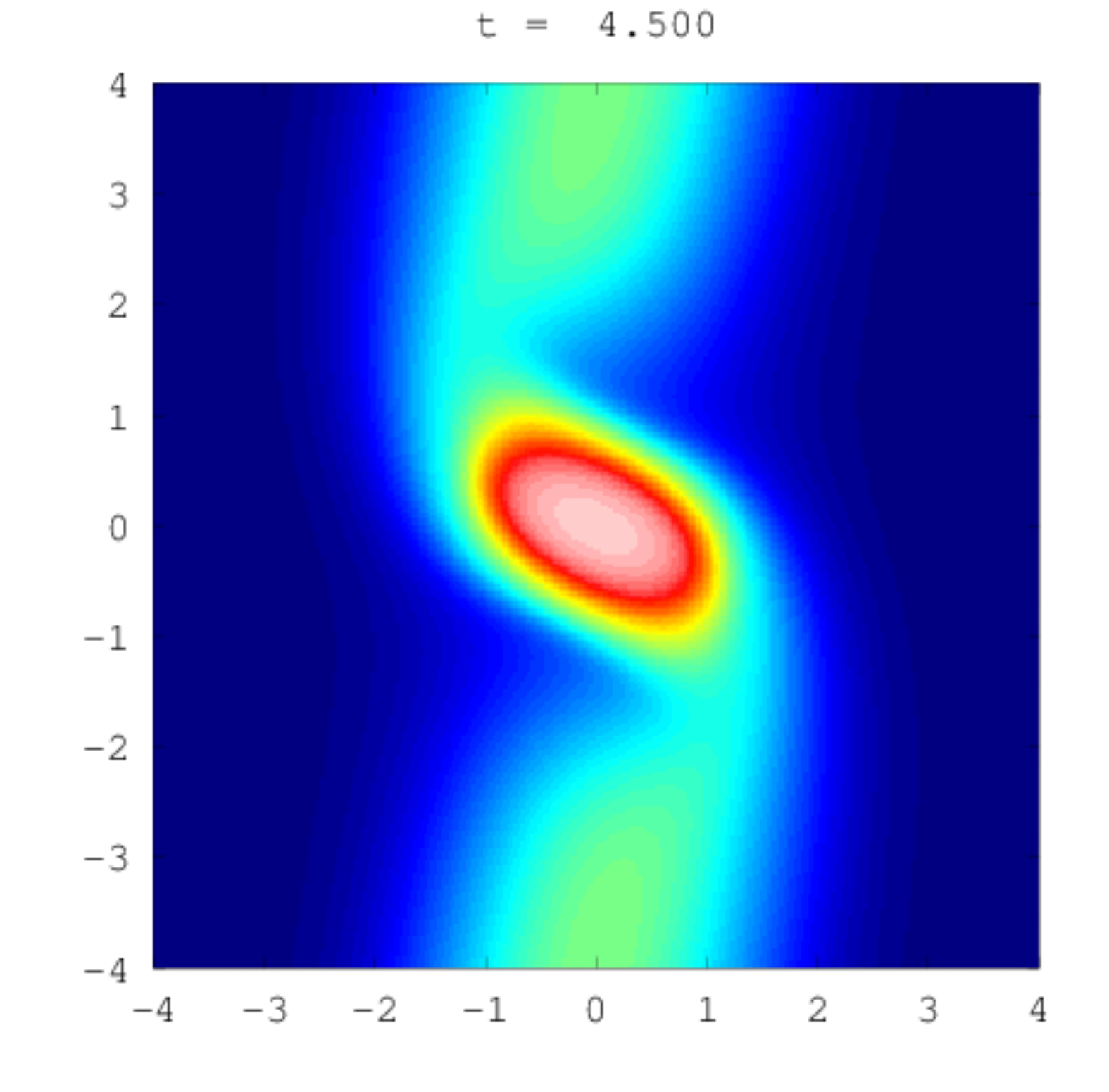}
\label{fig:MSA10}
}
\subfigure[{\footnotesize \  $t=80$.}]{\includegraphics[width=0.23\textwidth]{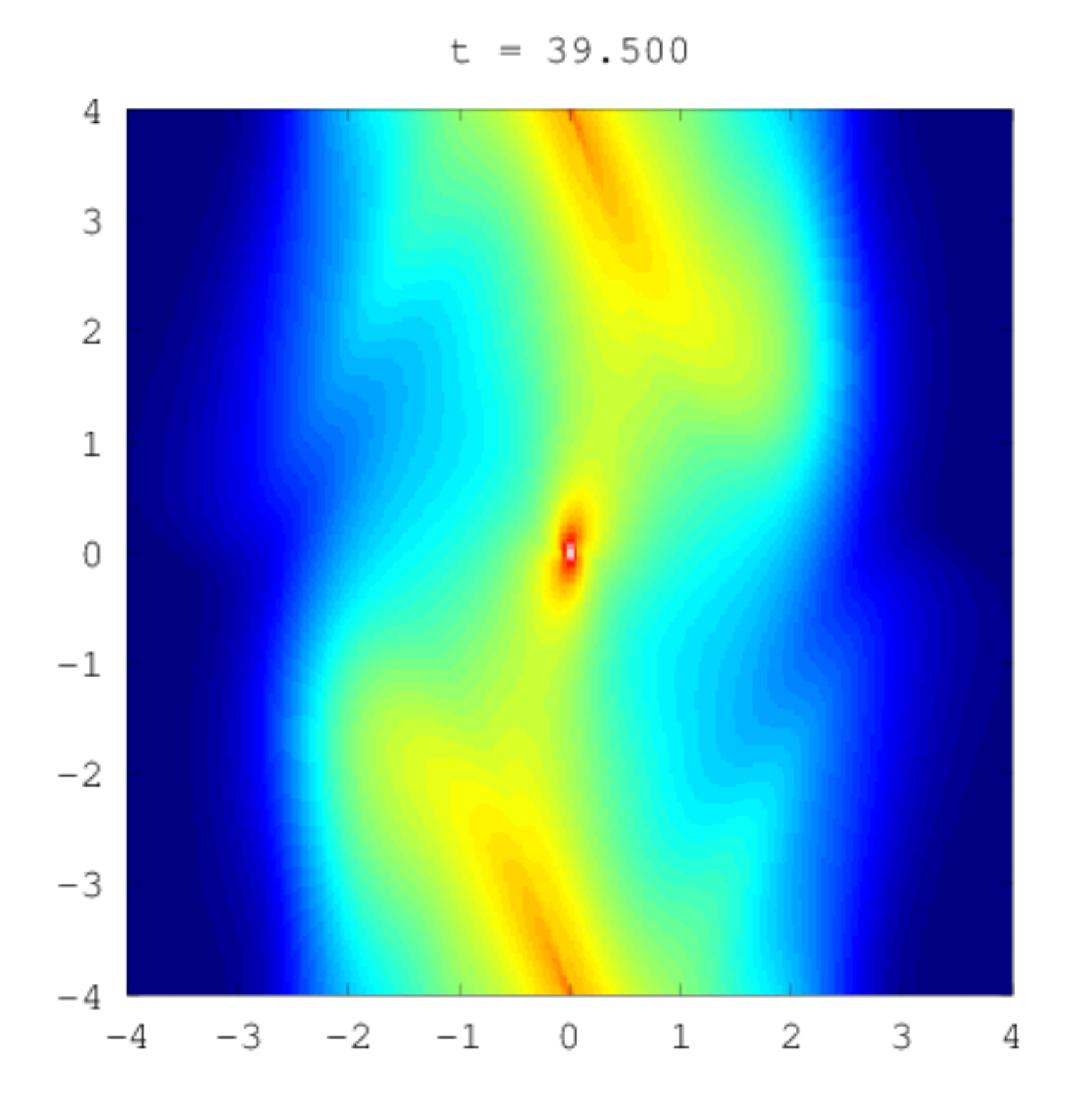}
\label{fig:MSA80}
}
\subfigure[{\footnotesize \  $t=500$.}]{\includegraphics[width=0.23\textwidth]{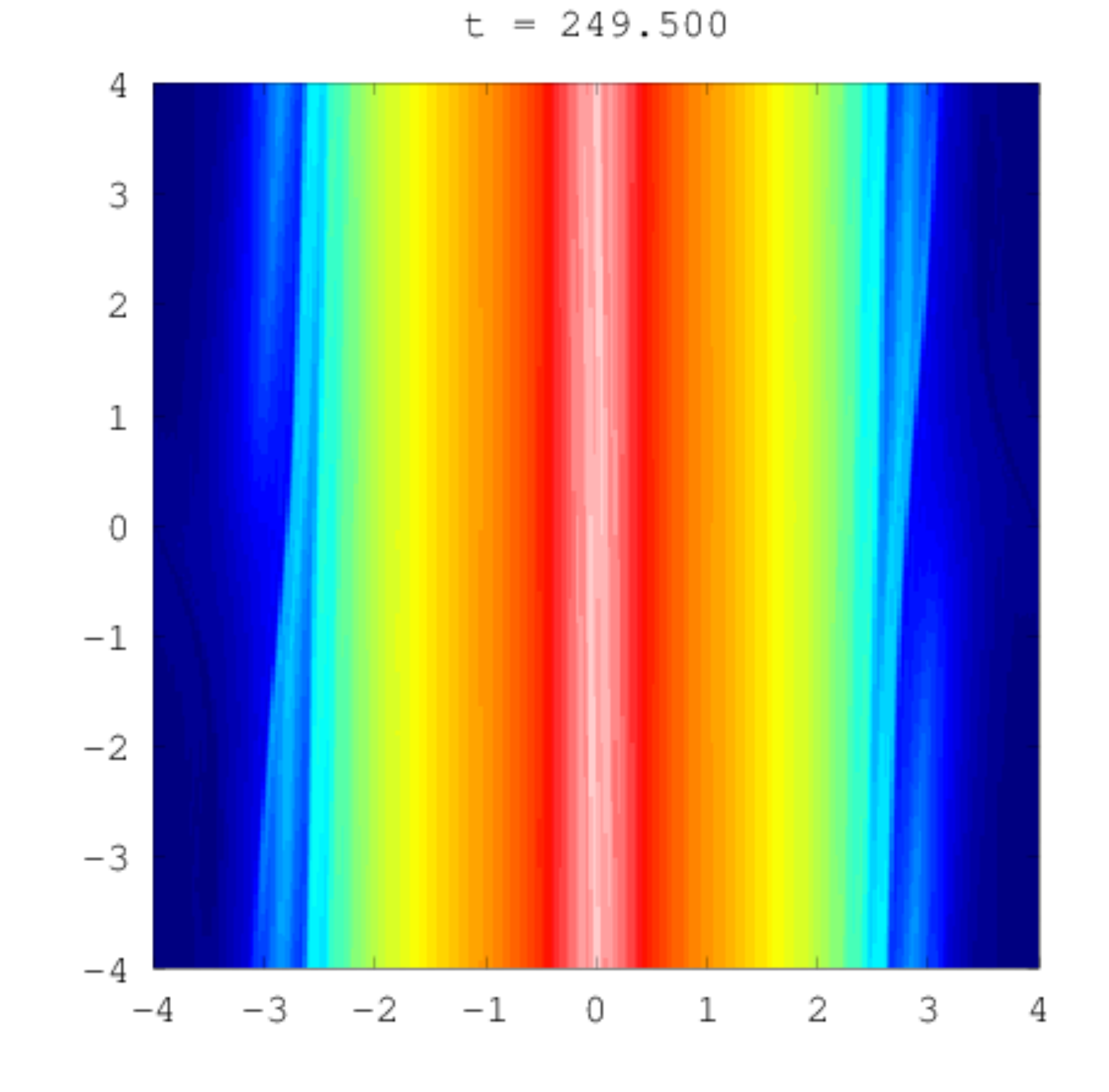}
\label{fig:MSA500}
}
\caption[sm]
{Shading  of potential vorticity under metriplectic evolution for (artificial) times.
}
\label{fig:MSA}
\end{figure}

As an example we consider a ridge described by $T=e^{-x^2/2}$ and integrate  forward using a pseudospectral code until the system approaches a relaxed state.   Figure \ref{fig:MSA} shows an initial dumbbell shaped concentration of vorticity being sheared out along the ridge as the artificial time (determined for convenience by the parameter $\ga$) progresses.

\begin{figure}[htb]
\includegraphics[width=0.4\textwidth]{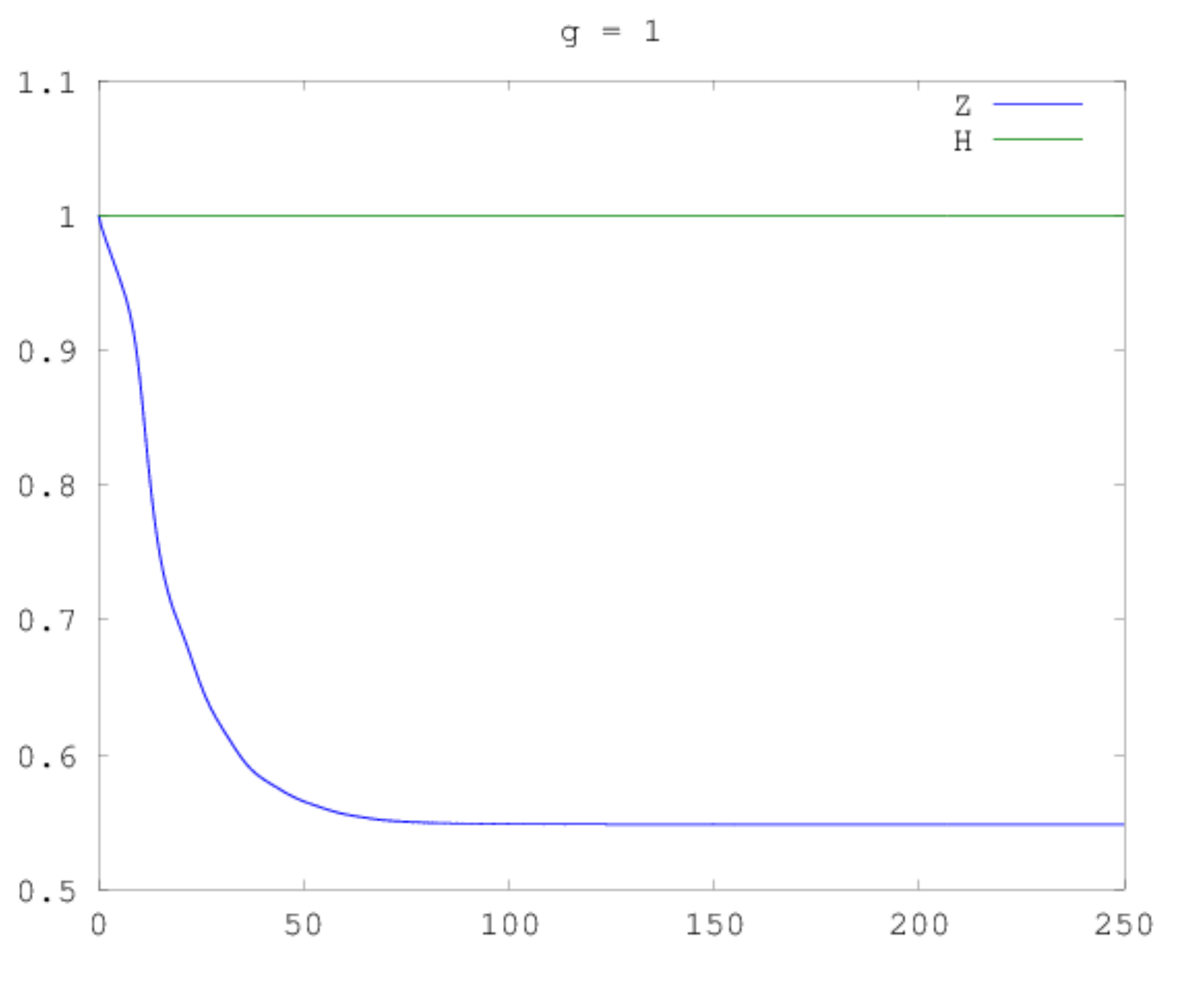}
\caption{Plot of energy and enstrophy under metriplectic evolution, with energy being conserved and enstrophy being decreased by the dynamics.}
\label{fig:qpsi}
\end{figure}

\subsubsection{Double Bracket SA}
\label{sssec:DBSA} 

\begin{figure}[htb]
\centering
\subfigure[{\footnotesize \ $t=0$.}]{
\includegraphics[width=0.2\textwidth]{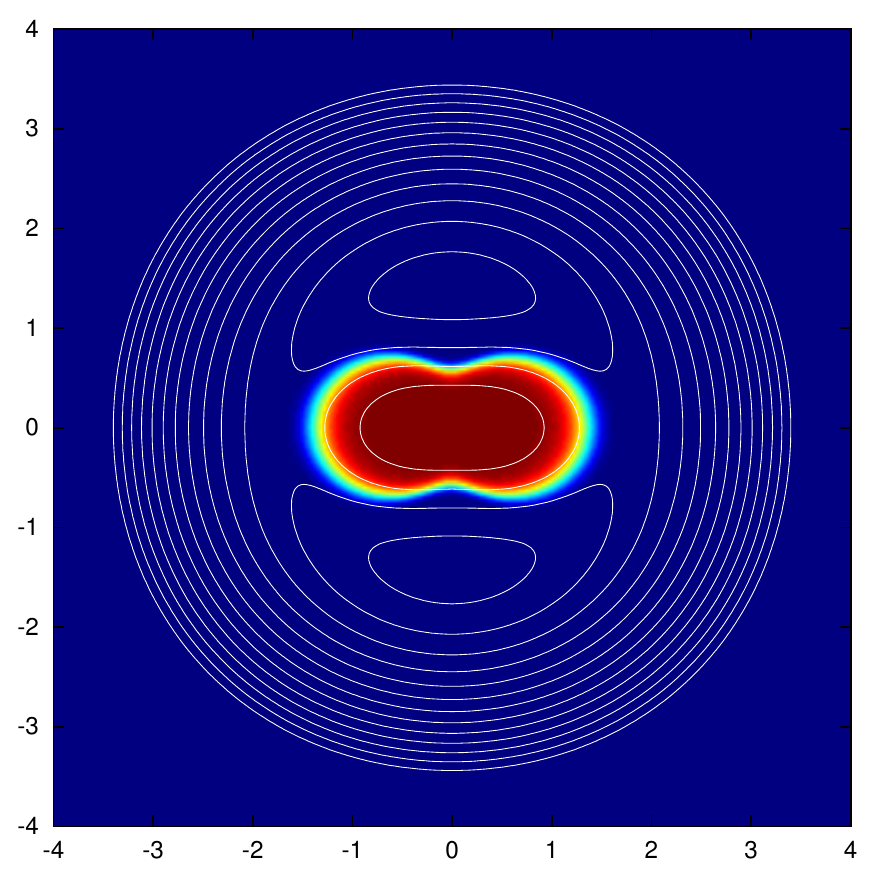}
\label{fig:els_4_m1_t04}
}
\subfigure[{\footnotesize  \ $t=80$.}]{
\includegraphics[width=0.2\textwidth]{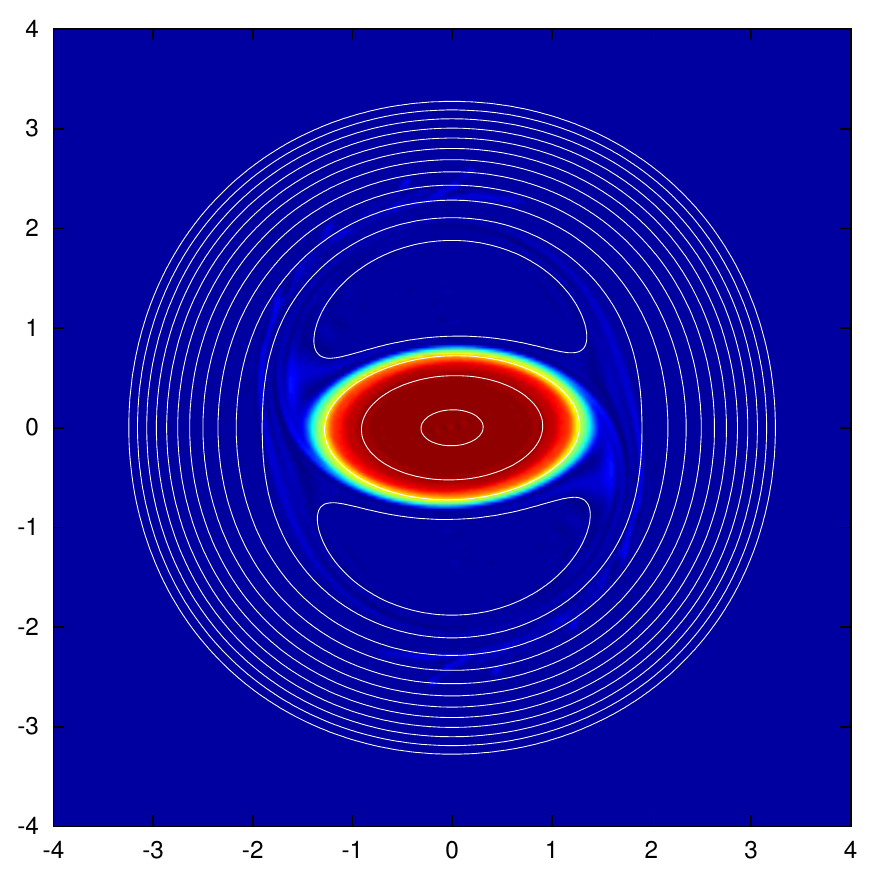}
\label{fig:els_4_m1_t08}
}
\subfigure[{\footnotesize  \ $t=120$.}]{
\includegraphics[width=0.2\textwidth]{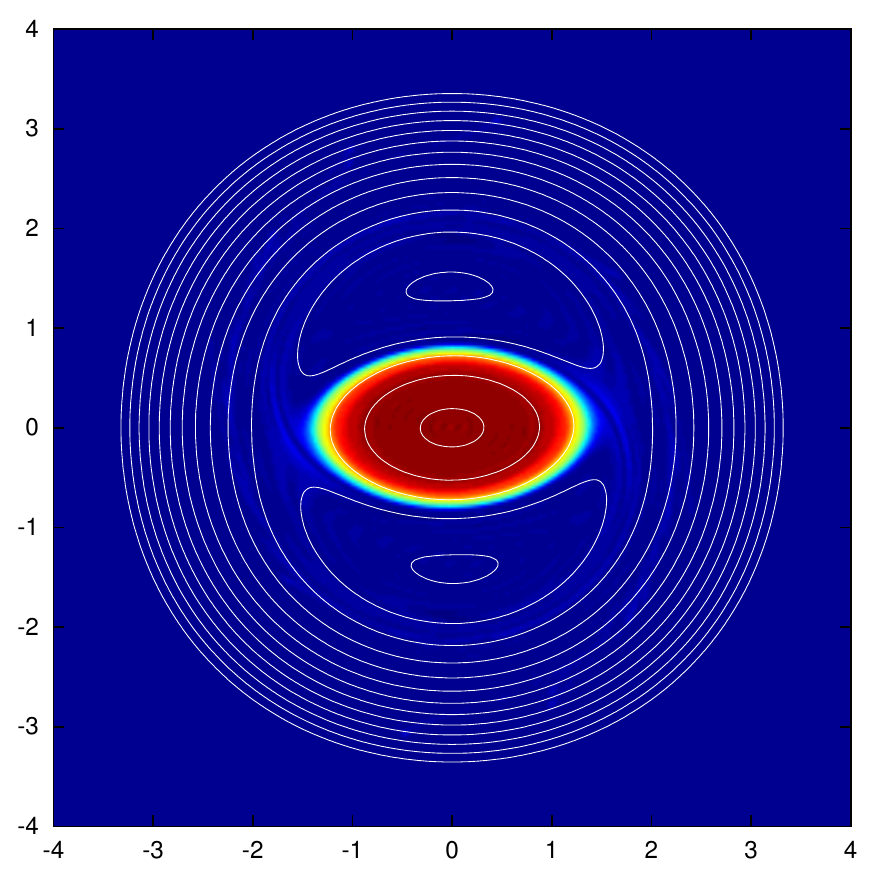}
\label{fig:els_4_m1_t12}
}
\subfigure[{\footnotesize  \ $t=200$.}]{
\includegraphics[width=0.2\textwidth]{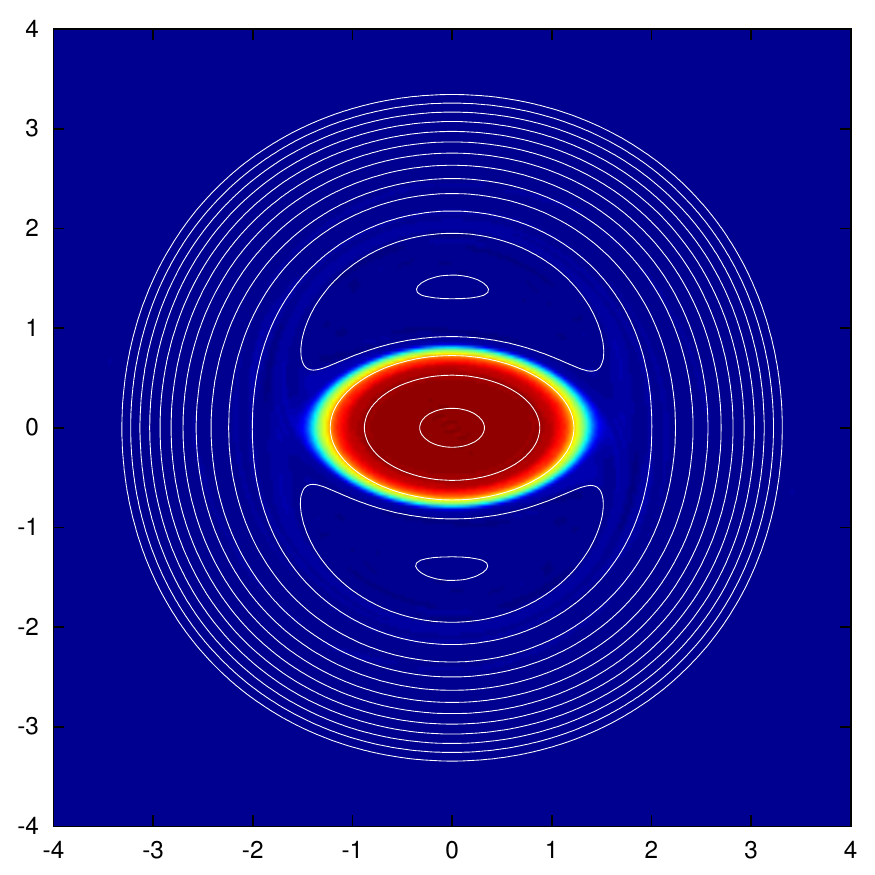}
\label{fig:els_4_m1_t20}
}
\caption[ellipse 1 under DSA-dynamics $\si=1$]
{Vorticity (shading) and stream function (contours) under   double bracket SA with a  two-fold symmetric initial  condition.  The code strives to maximize the energy subject to the  enforcement of  Dirac constraints that select out the equilibrium.  The dynamics effects  a symplectic rearrangement of the initial condition.}
\label{fig:els_4_m1}
\end{figure}

The original double bracket formulation of Ref.~\onlinecite{vallis89} was generalized in two  ways in Ref.~\onlinecite{pjmF11}:  by introducing a smoothing kernel and by imposing constraints using Dirac brackets.  This allowed a much wider class of states to be found.  In subsequent work this formalism has been applied to reduced MHD.\cite{furukawa1,furukawa2,pjmF16,pjmBKMS16}  Here, in Fig.~\ref{fig:els_4_m1},  we display a result from  Ref.~\onlinecite{pjmF11}  for Euler's equation for ideal  vorticity dynamics in two dimensions.   In this example  we seek a rotating state for the scalar vorticity that has two-fold symmetry.   Double bracket SA is used to find this as  an equilibrium in a rotating  frame of reference.  We refer the reader to  Ref.~\onlinecite{pjmF11} for details and many examples. 
  
  \section{GEMPIC: Geometric ElectroMagnetic Particle-In-Cell Methods}
  \label{sec:gempic}

  The results of this section are built on the flurry of  discoveries by  Evstatiev and 
  Shadwick\cite{Evstatiev:2010,Evstatiev:2013,Evstatiev:2014,Evstatiev:2014,Evstatiev:2014b}, the  voluminous work of Qin and collaborators,\cite{Squire:2012,xiao13,He:2015,Xiao:2015,qin16,He:2016,pjmXQLYZH16}  and the earlier work of Sonnendr\"ucker  and collaborators of a geometrical nature (e.g.\ Ref.~\onlinecite{Back:2014}) culminating  in the GEMPIC code described  by  Kraus et al.\  in  Ref.~\onlinecite{gempic}.  The essence of the GEMPIC procedure was contained in the  presentations at  this meeting by Qin\cite{hongAPS} and Kraus.\cite{pjmKKS16b}   GEMPIC is a functioning Maxwell-Vlasov PIC code that has been tested  and  is available in the  SelabLib library.\cite{selalib}  Many additional references to the relevant  literature can be found  in Ref.~\onlinecite{gempic}.

 GEMPIC is a  culmination that uses  many of the ideas about structure presented in previous sections of this paper. A semidiscrete  reduction of the full Maxwell-Vlasov system is obtained,  either from its variational principle  or by reduction of its noncanonical Poisson bracket, resulting in a finite-dimensional noncanonical Hamiltonian theory.  The finite-dimensional theory exactly  conserves invariants, including finite-dimensional versions of the  energy and Casimir invariants, and a symplectic method, specifically a  PI splitting method,  is devised for its numerical solution.    Thus GEMPIC brings together many ideas presented here,  but it  also uses ideas for the solution of Maxwell's equation based on its geometric structure.  
 
 In Sec.~\ref{ssec:MVhamS} we review the HAP structure of the Maxwell-Vlasov system, while in Sec.~\ref{ssec:DHAP} we describe its semidiscretization and resultant Hamiltonian structure. Obtaining this structure is complicated, so  only a sketch is given,  with a goal of giving the  flavor of the calculations of Ref.~\onlinecite{gempic}.  The section ends with  Sec.\ref{ssec:split},   a discussion of  the splitting method.

  \subsection{Maxwell-Vlasov  HAP Structure}
  \label{ssec:MVhamS}

An action principle for  the  Maxwell-Vlasov system was proposed by Low,\cite{Low:1958} which can be viewed as a continuum or smoothed out version of the nonrelativistic limit of the discrete particle action principle of  \eqref{kinetic}--\eqref{field}.  Low's action treats the particles by means of a Lagrangian or material continuum variable labeled by its  initial condition, which  replaces  the discrete particle index $i$.   For simplicity we suppress entirely  the sum over species with its presence being implicit.   Additional material on action principles for the  Maxwell-Vlasov system can be found  in Refs.~\onlinecite{dewar72,pjmY92}.

Suppose at $t=0$  a `particle'  is located at every point $\mathring{z}:=(\ringx,\ringv)$
of phase space.  Thus, in the action of \eqref{kinetic}--\eqref{field} we replace the discrete
label by this continuum label,  $\bfq_{i}\rightarrow \bfq(\ringz,t)$, and we replace 
the sums as follows:
\[
 \sum_{i=1}^{N}\rightarrow \int\! d\mathring{z} \, f_{0}(\ringz)\,,
 \]
where $d\mathring{z}=d^3\ringx\,d^3\ringv$ and 
$f_{0}$ can be viewed as a probability or number density attached to each point 
of phase space.  This procedure results in the following action:  
\bqy
S[q,\phi,A]  &=&   \int \!dt \!\int \!d\ringz\, f_{0}(\ringz)\, \frac{m}{2}\,
       |\dot{\bfq}(\ringz,t)|^{2}
       \label{low}\\
 &-& e \int \!dt\! \int \! d\ringz \,  f_{0}(\ringz) \int \!d^3x\,
      \Big[\phi(\bfx,t)
      \nonumber\\
      && \hspace{1.5 cm}  - \frac{\dot{\bfq}}{c} \cdot \bfA(\bfx,t) \Big]\,
      \delta\bigl(\bfx-\bfq(\ringz,t)\big)
      \nonumber\\
  &+&    \frac1{8\pi}\int \! dt \!\int\! d^3x\,  \Big[|\bfE(\bfx,t)|^{2}
- |\bfB(\bfx,t)|^{2}\Big]  \, ,  
\nonumber
\eqy
where now all variables are fields, the particle phase space field $\bfq(\ringz,t)$,
and the electromagnetic fields $\phi(\bfx,t)$ and $\bfA(\bfx,t)$.

The functional derivative $\delta S[\bfq,\phi,\bfA]/\delta
\bfq(\ringz,t)$ produces the partial differential equation for Lagrangian particle orbits.  This equation  can be
shown to be equivalent to the Vlasov equation if we define  the usual Eulerian expression for the distribution function    as 
 $f(\bfx,\bfv,t):= f_{0}(\ringz)$,  where $\ringz$ is the initial condition of the
particle located at $z$ at time $t$, i.e., $\ringz$ is determined by
inverting $z=(\bfx,\bfv)=(\bfq(\ringz,t),\dot{\bfq}(\ringz,t))$.  This inversion of  the map $\ringz\leftrightarrow z$ is  possible  by uniqueness  and, moreover,   it  has unit Jacobian.  

 As with the discrete particle action of \eqref{kinetic}--\eqref{field}, Faraday's law and $\nabla\cdot \bfB=0$ follow because of  the introduction of the potentials, $\phi$ and $\bfA$ with the relations \eqref{potentials}, and the remaining two Maxwell equations  follow from $\de S/\de \phi(\bfx,t)$ and $\de S/\de \bfA(\bfx,t)$, but now with the sources
\bqy
\rho(\bfx,t)&=& \sum_s e\int\!d^3v\, f(\bfx,\bfv,t)
\label{VMcharge}
\\
 \bfJ(\bfx,t)&=&  \sum_s e\int\!d^3v\, f(\bfx,\bfv,t) \, \bfv\,.
\label{VMcurrent}
\eqy

Thus with the interpretation above, Low's action has the Maxwell-Vlasov system as its  Euler-Lagrange equations, which for completeness we record below, 
\bqy
   \frac{\p f}{\p t}&=&  -   \mathbf{v}\cdot \nabla  {f}- \frac{e}{m}\,\left(\mathbf{E} +\frac{\mathbf{v}}{c}\times \mathbf{B}\right)\cdot \frac{\p   f}{\p  \mathbf{v}}\,,
   \label{usualVM}
   \\
 \frac{\p \bfE}{\p t} &=&\nabla\times\bfB -4\pi \bfJ \,,
 \label{dotE2}
 \\
 \frac{\p \bfB}{\p t} &=&-\nabla\times\bfE\,,
  \label{dotB}
 \\
&&\hspace{-1 cm} \nabla\cdot\bfB =0\, \qquad \mathrm{and}\qquad \nabla\cdot\bfE=4\pi \rho  \,.
\label{constraints}
\eqy

Note that \eqref{low} reverts to \eqref{kinetic}--\eqref{field} if we suppose that initially particles are located at $N$
isolated points, i.e.\ upon using 
\bq
f_{0}(\ringz)=\sum_{i=1}^{N} w_i \delta(\ringz- \ringz_{i})\,,
\eq
 $\bfq^{i}(t):= \bfq(\ringz_{i},t)$,   and setting  all the `weights' $w_i=1$.

The counterpart of the Low action is the purely Eulerian gauge-free noncanonical  Hamiltonian theory  conceived of in Ref.~\onlinecite{pjm80} that uses the natural field variables $(f,\bfE,\bfB)$.  This theory uses the conserved Maxwell-Vlasov-energy as its Hamiltonian,    
\bq
H=\frac{m}{2}\int \! d^3x\, d^3v \, |\bfv|^2 \, f + \frac1{8\pi} \int \!d^3x \big(|\bfE|^2 + |\bfB|^2\big)\,, 
\label{VMham}
\eq
together with the following noncanonical Poisson  bracket
\bqy
\Brac FG&=&\frac1{m}\int \! d^3x \,d^3v \,\Big(  f\,  \brac{F_f}{G_f}
 \label{MVbkt}\\ 
&+&\frac{e}{m c} \, f  \, \bfB\cdot\big(\p_{\bfv}F_f \times \p_{\bfv}G_f \big)
\nonumber\\
&+&  {4 \pi e}{}\, f \, \big(G_{\bfE}\cdot\p_{\bfv}F_f - F_\bfE \cdot\p_{\bfv} G_f\big)\Big) 
\nonumber \\
&+& 4\pi c \int\!d^3x \, \big(F_{\bfE}\cdot \nabla\times G_{\bfB} - G_{\bfE}\cdot\nabla\times F_{\bfB}\big)\,,  
\nonumber
\eqy
where $[f,g]=\nabla f\cdot  \p_{\bfv}g-\nabla g\cdot  \p_{\bfv}f$.    With the  noncanonical  bracket of \eqref{MVbkt} and the Hamiltonian  \eqref{VMham},  one obtains the Vlasov-Maxwell system of \eqref{usualVM}, \eqref{dotE2}, and \eqref{dotB} as 
 \bq
  \frac{\p f}{\p t}= \{f,H\}\,,  \quad \frac{\p \bfE}{\p t}= \{\bfE,H\}\,,  \quad \frac{\p \bfB}{\p t}= \{\bfB,H\} \,.
\eq

The bracket of \eqref{MVbkt} was  introduced in Ref,~\onlinecite{pjm80}, with a term  corrected in Ref.~\onlinecite{Marsden:1982} (see also Ref.~\onlinecite{pjm82}), and its limitation to divergence-free magnetic fields first pointed out in Ref.~\onlinecite{pjm82}. See also Refs.~\onlinecite{pjmCGBT13} and \onlinecite{pjm13}, where the latter contains the details of the direct proof of the Jacobi identity,  
\bqy
&& \{\{F,G\},H\} + \mathrm{cyc} =
\label{MVjac}\\
&&\hspace{1.5 cm}\frac{e}{m^3 c}\int \! d^3x \,d^3v \, f\, \nabla\cdot \bfB \,  
\big(\p_{\bfv}F_f\times \p_{\bfv}G_f\big)\cdot \p_{\bfv}H_f\,.
\nonumber
\eqy
Thus the Jacobi identity is satisfied for arbitrary functionals $F,G,H$ defined on divergence-free magnetic fields.  The constraints of \eqref{constraints}, 
\bq
 C_B: = \nabla\cdot\bfB =0  \quad \mathrm{and}\quad C_E:= \nabla\cdot\bfE-4\pi \rho  
\eq
satisfy   $\{F,C\}=0$ pointwise for all $F$.  Thus, $C_E$ is a Casimir invariant, while $C_B$ could be called a semi-Casimir, because it is intertwined with the satisfaction of the Jacobi identity per \eqref{MVjac}.

The relativistic Vlasov-Maxwell theory follows upon  replacing $m|\bfv|^2/2$ in the first term of  the Hamiltonian \eqref{VMham} with  $-mc^2\sqrt{1 - |\bfv|^2/c^2}$,  and the theory can be written in manifestly covariant form,\cite{pjmP85,pjmMMT86} but  this will not be pursued  further here.

 \subsection{Discrete HAP Maxwell-Vlasov}
 \label{ssec:DHAP}

Let us now consider semidiscrete reductions  of the Maxwell-Vlasov system, systems of ODEs that are designed to approximate the dynamics.  Many approaches for this have been pursued, but that most natural to a physicist is to start from the action principle,  because this provides a direct avenue for maintaining the  exact conservation of energy and other invariants in the reduced theory.   Many such approximations have been made in plasma physics, e.g.\ that of Refs.~\onlinecite{dewar72,pjmEH03,pjmEH05},  where the latter two were specifically  designed for  obtaining  computable models,   the  single- and multi-wave models of plasma physics.\cite{pjmBT13}  However,  specifically for the purpose of finding a semidiscrete reduction for solving  the Maxwell-Vlasov system,  the early work of  Ralph Lewis on variational PIC stands out.\cite{lewis70,lewis70b,lewis72}  His approach to PIC did not take hold, most likely because NGP (nearest-grid-point) was used for the force which proved to be prohibitively  noisy for computers of the day. 

The idea of Lewis  was to insert expansions for the fields and particles in terms of  sums over  basis functions into \eqref{low}, truncate the sums, integrate over $\mathring{z}$, and then vary as in Hamilton's principle of mechanics to obtain sets of ODEs that exactly conserve  invariants.  The ODEs obtained by this procedure describe both particle and field degrees of freedom with their coupling provided automatically by the coupling terms of  \eqref{low}.   This variational approach of Lewis was revived in more recent works,\cite{Evstatiev:2010,Squire:2012,Evstatiev:2013,xiao13}  where the use of better basis functions and/or the introduction of shape functions was seen to dramatically reduce  noise and other sources of error.  For example, upon inserting  Fourier series  for the potentials, 
\bq
\phi= \sum_{\bfk} \phi_{\bfk} \,e^{i\bfk\cdot\bfx} \quad \mathrm{and}\quad \bfA=\sum_{\bfk} \bfa_{\bfk} \,e^{i\bfk\cdot\bfx}\,,
\label{fourier}
\eq
into the field term  of \eqref{low} and representing the particles  by a  shape function $\cals$  as follows:
\bq
f(\bfx,\bfv,t)=\sum_i\,  w_i\,\cals (\bfx-\bfq_i(t))\de (\bfv-\dot{\bfq}_i(t))
\label{smooth}
\eq
 where $\int d^3x \, \cals =1$, yields an action of the form
 \bq
 S[\bfq_i,\phi_{\bfk},\bfA_{\bfk}] = \int \!\!dt \, L(\bfq_i,\dot{\bfq}_i,\phi_{\bfk}, \bfA_{\bfk}, \dot{\bfA}_{\bfk})\,.
 \eq
Then,  extremization gives Lagrange's equations in terms of the Lagrangian $L$ in the usual manner.   By construction,  these equations  are conservative and have Hamiltonian structure. 

GEMPIC differs from the above procedure by using advances in the integration of Maxwell's equations that exploit  its intrinsic geometry, viz., that $\bfE$ is a 1-form and $\bfB$ is a 2-form.  Because of its  desirable properties,   the early finite difference discretization of Yee\cite{Yee:1966} was generalized to a complete discrete differential calculus  in later work,\cite{Desbrun:2008} and these ideas can be incorporated into a   Maxwell-Vlasov PIC code.  However, GEMPIC adapts   Finite Element Exterior Calculus (FEEC), a  finite element or spline approach that is described in detail in Refs.~\onlinecite{Arnold.Falk.Winther.2006.anum,Arnold.Falk.Winther.2010.bams}.   The upshot of this approach is that one obtains discrete versions of the grad, div, and curl operations, in terms of matrices, with properties like curl grad =0 maintained. An important aspect of this  approach is that there are now many finite element spaces, splines etc.\ that achieve this.   In fact,   the Fourier bases of \eqref{fourier} is a special case of the general framework. 

To see how this progresses, consider the expansions
\begin{align}\label{eq:fields_discrete}
\bfE_h = \sum_{i=1}^{N_1} e_i(t) \Lab_i^1, ~~~~~~~~
\bfB_h = \sum_{i=1}^{N_2} b_i(t) \Lab_i^2,
\end{align}
where $\Lab_i^{1,2}$ are the 1-form and 2-form bases with the desired properties.  Given these properties,  Eqs.~\eqref{potentials}
 take the discrete matrix forms
\begin{align}
\ce = - \G \cphi - \frac{\dd \ca}{\dd t}, ~~~~~~
\cb =   \C \ca\,, 
\end{align}
 where $\ce$ and $\cb$ have  as components the $e_i$ and $b_i$,  $\cphi$ and $\ca$ are the amplitudes for expansions analogous to \eqref{eq:fields_discrete} for   the potentials $\phi$ and $\bfA$, and $\G$ and $\C$ are the discrete gradient and curl operators.  The discrete form of Maxwell's equations become the  following ODEs in matrix form:
 \begin{align}
M_1 \frac{\dd \ce}{\dd t} - \C^\top M_2 \cb &=4\pi  \cj \label{systfe2damp} , \\
\frac{\dd \cb}{\dd t} +\C \ce &= 0 \label{systfe2dfar} , \\
\G^\top M_1 \ce &= 4\pi \crho  \label{systfe2dgauss} , \\
\D \cb &= 0 \label{systfe2dgaussm}\,, 
\end{align}
 where $\D$ is the discrete divergence operator, $\cj$ and $\crho$ are suitable $n$-tuples representing the charge and current densities  (3-forms and 2-forms) defined in terms of the particle degrees of freedom, $\top$ means transpose, and the matrices $M_{1,2}$ are defined by the pairing of the bases elements $ \Lab^{1,2}$ which in general are not orthogonal. 
 
 This discretization of Maxwell's equations is coupled with a particle discretization like \eqref{smooth} to obtain a finite-dimensional system.   That this resulting system possesses noncanonical Hamiltonian form, follows upon projecting the Poisson bracket of \eqref{MVbkt}.   This is done by using functional chain rule expressions such as those given in Refs.~\onlinecite{pjm81a,pjm81b}, e.g.,
 \bq
 \frac{\p \hat{F}}{\p \bfq_i}=\left. w_i \nabla \frac{\de F}{\de f}\right|_{(\bfq_i, \bfv_i)}
 \eq
 with $\bfv_i=\dot{\bfq}$ being the canonical momentum conjugate to $\bfq_i$.   To see the details of this  calculation we refer the reader to Ref.~\onlinecite{gempic}, but the final result is a Poisson bracket acting on  functions $\hat{F}$ and $\hat{G}$ with the arguments 
 \bq
 \mathbf{z}=  (\X, \V, \ce, \cb)
 \eq
 where $(\X, \V)$ denote the $6N$ particle degrees of freedom and  $\bfz$ in  index form,  $z^a$,  has  $a=1,2,\dots, M$ with $M$ including the particle and  $\ce$ and $\cb$ degrees of freedom.  Thus the Poisson bracket becomes 
\begin{align*}
\{\hat{F},\hat{G}\} = \frac{\p \hat{F}}{\p z^a} \jac^{ab} \, \frac{\p \hat{G}}{\p z^b} ,
\end{align*}
with the Poisson matrix  given by 
\bqy
\jac  &=&
\nonumber\\
&&\hspace{-.5cm} 
\begin{pmatrix}
0 &\hspace*{-.2cm} \MM_p^{-1} &\hspace*{-.2cm} 0 &\hspace*{-.2cm} 0 \\
- \MM_p^{-1} &\hspace*{-.2cm} \MM_p^{-1} \MM_q \mathbb{B}\, \MM_p^{-1} &\hspace*{-.2cm} \MM_p^{-1} \MM_q \LaB^1 M_1^{-1} &\hspace*{-.2cm} 0 \\
0 &\hspace*{-.2cm} - M_1^{-1} \LaB^{1\top} \MM_q \MM_p^{-1} &\hspace*{-.2cm} 0 &\hspace*{-.2cm}  M_{1}^{-1} \C^\top M_{2}^{-1} \\
0 &\hspace*{-.2cm} 0 &\hspace*{-.2cm} - M_{2}^{-1} \C M_{1}^{-1} &\hspace*{-.2cm} 0 \\
\end{pmatrix} .
\nonumber
\eqy
where 
\begin{align}
\MM_{p} = M_{p} \otimes \mathbb{I}_{3 \times 3} ,
\hspace{3em}
\MM_{q} = M_{q} \otimes \mathbb{I}_{3 \times 3} ,
\end{align}
with  $\mathbb{I}_{3 \times 3}$ denoting  the $3 \times 3$ identity matrix and 
$M_p$ and $M_q$ being  diagonal matrices with elements $m_i w_i$ and $e_i w_i$, respectively. The matrices $\mathbb{B}(\X, \cb)$
 and $\LaB^1(\X)$,   depend on components of $\bfz$ and  are given in terms of  $\Lab_i^{1,2,3}$ with $\Lab_i^{3}$ being the 3-form basis elements (see Ref.~\onlinecite{gempic} for details).  Although complicated, in the end $\jac$ is just a matrix of  functions of the dynamical variables, which was shown directly in Refs.~\onlinecite{gempic,He:2016} to satisfy the Jacobi identity, and thus with a Hamiltonian we have a finite, yet very large, noncanonical Hamiltonian system of the form of \eqref{NCham}.

 \subsection{Hamiltonian Splitting}
 \label{ssec:split}
 
 The discretized form of the Hamiltonian of \eqref{VMham} has three parts, $\hat{H} = \hat{H}_p +\hat{H}_E +\hat{H}_B$, with 
 \bq
 \hat{H}_p = \frac12 \V^\top \MM_p \V\,, \quad  \hat{H}_E = \frac12 \ce^\top \MM_1 \ce\,,
 \quad \hat{H}_B = \frac12 \cb^\top \MM_2 \cb\,.
 \nonumber
 \eq
 This decomposition is the basis of the splitting method developed in  Refs.~\onlinecite{Crouseilles:2015,Qin:2015,He:2015}.  The essential idea is that  each of the subsystems
 \bq
 \dot\bfz=\{\bfz,  \hat{H}_p \}\,,\quad  \dot\bfz=\{\bfz,  \hat{H}_E \}\,,  \quad\dot\bfz=\{\bfz,  \hat{H}_B \}\,,
 \eq
 can be solved exactly and each is a Poisson map.  Thus, their composition is a Poisson map, and we can construct a time step that exactly conserves the Poisson bracket.  This is an example of the PI discussed in Sec.~\ref{ssec:PI}.  
 
 The integrator is constructed as a Lie series as in \eqref{lie}.  For example a first order integrator is given by 
\begin{align}
\bfz (\Dt)
= e^{ \Dt \, X_{E}  }
   e^{\Dt \, X_{B}  }
   e^{\Dt \, X_{p_1}}
e^{ \Dt \, X_{p_2} }
  \, \bfz(0) ,
\end{align}
where  the $X$s are Hamiltonian vector fields, e.g., 
\[
X^a_{E}= \jac^{ba} \frac{\p \hat{H}_E}{\p z^b}\,.
\]
For different arrangements of exponentials, one can obtain a second order integrator --  we direct the reader to the references for  details. 
  
  \section{Conclusions} 
  \label{sec:conclu}

 In this paper  HAP formulations of  plasma dynamical systems have been reviewed, and methods for integrating differential equations that preserve  such structure have been surveyed.  In Sec.~\ref{sec:CI}   CI schemes that  exactly conserve constants of motion were discussed.  Then in Sec.~\ref{sec:SI}  SI, which   preserves  canonical Hamiltonian structure, viz.\ the areas of  \eqref{sympA}, and the associated  Poincar\'e invariants  and   loop integrals of \eqref{loop}, was described.   This was followed by two other kinds of HI,  the VI of Sec.~\ref{ssec:VI},  based on approximating variational principles, and the PI of Sec.~\ref{ssec:PI} for Hamiltonian systems with the  noncanonical Poisson bracket formulation that is typical of plasma models.  Next MI was proposed in  Sec.~\ref{sec:MI}, an integration scheme   that would preserve the structure of systems that are conservative with  both Hamiltonian and dissipative parts.   Simulated annealing, the relaxation method for obtaining equilibrium states was described  in Sec.~\ref{sec:SA}.  Examples of metriplectic SA and double bracket SA were given.  Finally,  in Sec.~\ref{sec:gempic} the  GEMPIC algorithm for the Maxwell-Vlasov (MV) system are described.  The GEMPIC code is a fitting finale, for it is  a  culmination of the work of many researchers that exactly  preserves geometric structure described in previous sections of this  paper.   
 
Given the many time-honored computational methods, it is important and obvious to note that structure preserving methods like those described here may be unnecessary.  However, for some problems it may be easy and inexpensive to try one of the methods described here,  and   depending on the problem being addressed it  might prove to be superior,  but no structure preserving integrator is going to be a  panacea for all computational ills.   For some problems such techniques are useful, and  may even be  essential to obtain an accurate solution.

There are many avenues for future work,  both of a general nature and specific to   plasma physics.    Since suggesting the idea of MI, several researchers have already taken up the challenge and have made  progress.  It appears in the PDE context that MI will be useful for handling  collision operators.  Application to drift kinetic or gyrokinetic theories is underway\cite{Evstatiev:2014c}, but  because of the different variational and Hamiltonian structure of these theories (see e.g.\ Refs.~\onlinecite{pjmBBQ15,pjmBBGV16}),  modifications are necessary.  Lastly we  mention that   Hamiltonian reductions like that of GEMPIC and others\cite{burby16}
open the field of {\it discrete gyrokinetics}.  Conventional gyrokinetics reduces an  infinite-dimensional theory, the MV system,  to another reduced  infinite-dimensional theory.  Given a discretization like GEMPIC one could proceed in two novel ways: i) start from the finite-dimensional Hamiltonian theory of the discretization and then use  the considerable lore of more than a century of rigorous Hamiltonian perturbation methods to arrive  at a reduced faithful system.   This is facilitated by the adaptation of canonical perturbations methods to noncanonical\cite{pjmV16,pjmVC16} or ii) use multiscale methods (e.g.\ Ref.~\onlinecite{EBE03}) on the large but finite GEMPIC-like system, thereby circumventing  the usual gyrokinetic expansions by incorporating modern day numerical tools.  Neither of these approaches are trivial, and their efficacy remains to be determined.


\section*{Acknowledgment}
\noindent The author received support from the  U.S. Dept.\ of Energy Contract \# DE-FG05-80ET-53088 and from a Forschungspreis from  the Alexander von Humboldt Foundation.   He would like to  warmly  acknowledge the hospitality of the Numerical Plasma Physics Division of the IPP, Max Planck, Garching and enlightening conversations and/or correspondence with  many colleagues and friends, H.~Qin, B.~Shadwick, E.~Evstatiev, E.~Sonnendr\"ucker, M.~Kraus, J.~Burby,  G.~Flierl, J.~Finn,  L.~Chacon, I.~Gamba, Y.~Cheng, B.~Scott, O.~Maj, G.~Hammett, A.~Brizard, J.~Meiss, J.~Cary, F.~Pegoraro, T.~Antonsen, Y.~Zhou, Lee Ellison,  and others he has probably forgotten.







%


\end{document}